\documentclass[letterpaper]{aa}
\usepackage{txfonts}
\usepackage{graphicx}
\begin{document}

\title{
First Detection of Polarization of the Submillimetre Diffuse Galactic
Dust Emission by Archeops}

   \subtitle{}
\author{ 
A.~Beno{\^\i}t\inst{1} \and 
P.~Ade \inst{2} \and 
A.~Amblard \inst{3} \and 
R.~Ansari \inst{4} \and 
{{\'E}}.~Aubourg \inst{5, \, 6} \and
S.~Bargot \inst{4} \and
J.~G.~Bartlett \inst{7, \, 6} \and 
J.--Ph.~Bernard \inst{8} \and 
R.~S.~Bhatia \inst{9} \and 
A.~Blanchard\inst{10} \and 
J.~J.~Bock \inst{11,\,12} \and
A.~Boscaleri \inst{13} \and 
F.~R.~Bouchet \inst{14} \and 
A.~Bourrachot \inst{4} \and 
P.~Camus \inst{1} \and 
F.~Couchot \inst{4} \and
P.~de~Bernardis \inst{15} \and 
J.~Delabrouille \inst{7, \, 6} \and
F.--X.~D{\'e}sert \inst{16} \and 
O.~Dor{{\'e} \inst{17}} \and 
M.~Douspis \inst{18} \and 
L.~Dumoulin \inst{19} \and 
X.~Dupac \inst{9} \and
P.~Filliatre \inst{7, \, 6} \and 
P.~Fosalba \inst{14} \and
K.~Ganga \inst{20} \and 
F.~Gannaway \inst{2} \and 
B.~Gautier \inst{1} \and 
M.~Giard \inst{8} \and
Y.~Giraud--H{\'e}raud \inst{7, \, 6} \and 
R.~Gispert \inst{21\dag}\thanks{Richard Gispert passed away few weeks
after his return from the early mission to Trapani} \and 
L.~Guglielmi \inst{7, \, 6} \and
J.--Ch.~Hamilton \inst{22} \and 
S.~Hanany \inst{23} \and
S.~Henrot--Versill{\'e} \inst{4} \and 
J.~Kaplan \inst{7, \, 6} \and
G.~Lagache \inst{21} \and 
J.--M.~Lamarre \inst{24} \and 
A.~E.~Lange \inst{11} \and 
J.~F.~Mac{\'\i}as--P{\'e}rez \inst{25} \and 
K.~Madet \inst{1} \and 
B.~Maffei \inst{2} \and
Ch.~Magneville \inst{5, \, 6} \and
D.~P.~Marrone \inst{23} \and
S.~Masi \inst{15} \and 
F.~Mayet \inst{5} \and 
A.~Murphy \inst{26} \and
F.~Naraghi \inst{25} \and 
F.~Nati \inst{15} \and
G.~Patanchon \inst{7, \, 6} \and
G.~Perrin \inst{25} \and 
M.~Piat \inst{21} \and 
N.~Ponthieu \inst{25} \and
S.~Prunet \inst{14} \and
J.--L.~Puget \inst{21} \and
C.~Renault \inst{25} \and 
C.~Rosset \inst{7, \, 6} \and
D.~Santos \inst{25} \and
A.~Starobinsky \inst{27} \and
I.~Strukov \inst{28} \and
R.~V.~Sudiwala \inst{2} \and 
R.~Teyssier \inst{14, \, 29} \and
M.~Tristram \inst{25} \and
C.~Tucker\inst{2} \and
J.--C.~Vanel \inst{30} \and 
D.~Vibert \inst{14} \and 
E.~Wakui \inst{2} \and 
D.~Yvon \inst{5, \, 6}
}

   \offprints{benoit@grenoble.cnrs.fr}
   \mail{benoit@grenoble.cnrs.fr}

\institute{
Centre de Recherche sur les Tr{\`e}s Basses Temp{\'e}ratures,
BP166, 38042 Grenoble Cedex 9, France
\and
Cardiff University, Physics Department, PO Box 913, 5, The Parade,   
Cardiff, CF24 3YB, UK\and
University of California, Berkeley, Dept. of Astronomy, 601
Campbell Hall, Berkeley, CA 94720-3411, U.S.A.
\and
Laboratoire de l'Acc{\'e}l{\'e}rateur Lin{\'e}aire, BP~34, Campus
Orsay, 91898 Orsay Cedex, France
\and
CEA-CE Saclay, DAPNIA, Service de Physique des Particules, 
Bat 141, F-91191 Gif sur Yvette Cedex, France
\and
F{\'e}d{\'e}ration de Recherche APC, Universit{\'e} Paris 7, Paris, France
\and
Physique Corpusculaire et Cosmologie, Coll{\`e}ge de
France,  11 pl. M. Berthelot, F-75231 Paris Cedex 5, France
\and
Centre d'{\'E}tude Spatiale des Rayonnements,
BP 4346, 31028 Toulouse Cedex 4, France
\and
European Space Agency - ESTEC, Astrophysics Division, Keplerlaan 1, 2201 AZ Noordwijk, The Netherlands
\and
Laboratoire d'Astrophysique de l'Obs. Midi-Pyr{\'e}n{\'e}es,
14 Avenue E. Belin, 31400 Toulouse, France
\and
California Institute of Technology, 105-24 Caltech, 1201 East 
California Blvd, Pasadena CA 91125, USA
\and
Jet Propulsion Laboratory, 4800 Oak Grove Drive, Pasadena, 
California 91109, USA
\and
IROE--CNR, Via Panciatichi, 64, 50127 Firenze, Italy
\and
Institut d'Astrophysique de Paris, 98bis, Boulevard Arago, 75014 Paris,
France
\and
Gruppo di Cosmologia Sperimentale, Dipart. di Fisica, Univ.  ``La
Sapienza'', P. A. Moro, 2, 00185 Roma, Italy
\and
Laboratoire d'Astrophysique, Obs. de Grenoble, BP 53, 
 38041 Grenoble Cedex 9, France
\and
Department of Astrophysical Sciences, Peyton Hall - Ivy Lane, Princeton, 
NJ 08544-1001, USA
\and
Nuclear and Astrophysics Laboratory, Keble Road, Oxford,  OX1 3RH, UK
\and
CSNSM--IN2P3, B{\^a}t 108, 91405 Orsay Campus, France
\and
Infrared Processing and Analysis Center, Caltech, 770 South Wilson
Avenue, Pasadena, CA 91125, USA
\and
Institut d'Astrophysique Spatiale, B{\^a}t.  121, Universit{\'e} Paris
XI,
91405 Orsay Cedex, France
\and
LPNHE, Universités Paris VI et Paris VII, 4 place
Jussieu, Tour 33, 75252 Paris Cedex 05, France
\and
School of Physics and Astronomy, 116 Church St. S.E., University of
Minnesota, Minneapolis MN 55455, USA
\and
LERMA, Observatoire de Paris, 61 Av. de l'Observatoire, 75014 Paris, France
\and
Laboratoire de Physique Subatomique et de Cosmologie, 53 Avenue des Martyrs, 38026
Grenoble Cedex, France
\and
Experimental Physics, National University of Ireland, Maynooth, Ireland
\and
Landau Institute for Theoretical Physics, 119334 Moscow, Russia
\and
Space Research Institute, Profsoyuznaya St. 84/32, Moscow, Russia 
\and
CEA-CE Saclay, DAPNIA, Service d'Astrophysique, Bat 709, 
F-91191 Gif sur Yvette Cedex, France
\and
Laboratoire Leprince--Ringuet, Route de Saclay, 91128 Palaiseau Cedex
}

\date{\today}
   
\abstract{ We present the first determination of the Galactic
  polarized emission at 353~GHz by Archeops. The data were taken
  during the Arctic night of February 7, 2002 after the balloon--borne
  instrument was launched by CNES from the Swedish Esrange base near
  Kiruna. In addition to the 143~GHz and 217~GHz frequency bands dedicated to
  CMB studies, Archeops had one 545~GHz and six 353~GHz bolometers mounted in three
  polarization sensitive pairs that were used for Galactic foreground
  studies. We present maps of the $I,~Q,~U$ Stokes parameters over 17~\% of the
  sky and with a 13~arcmin resolution at 353~GHz
  ($850~\mu\mathrm{m}$). They show a significant Galactic large scale
  polarized emission coherent on the longitude ranges [100, 120] and
  [180, 200] deg. with a degree of polarization at the level of
  4--5~\%, in agreement with expectations from starlight polarization
  measurements. Some regions in the Galactic plane (Gem~OB1,
  Cassiopeia) show an even stronger degree of polarization in the
  range 10--20~\%. Those findings provide strong evidence for a
  powerful grain alignment mechanism throughout the interstellar
  medium and a coherent magnetic field coplanar to the Galactic
  plane. This magnetic field pervades even some dense clouds. Extrapolated
  to high Galactic latitude, these results indicate that interstellar dust
  polarized emission is the major foreground for PLANCK--HFI CMB
  polarization measurement.

         }

        \authorrunning{Beno\^\i t \& Archeops}
        \titlerunning{Polarization of diffuse Galactic Dust Emission}
   \maketitle


\section{Introduction}

The power spectrum of the {\sl temperature} anisotropies of the Cosmic
Microwave Background (CMB) have now been measured over most of the
relevant angular scales (10~arcmin to 90~deg, see a comparison of
different experiments in {\sl e.g.} \cite{benoit_cl} and
\cite{wmap_basic}). However, CMB {\sl polarization} is only in its
experimental infancy. Theoretical predictions are rather tight for the
polarization effect coming from the last scattering surface. Accurate
polarization measurements are not only useful for breaking some
degeneracies between cosmological parameters but also for obtaining
the gravitationnal wave background. Upper limits on polarization
(\cite{keating_01,oliveira}) are now superseded by detections by DASI
(\cite{dasi_pol}) and WMAP~(\cite{wmap_pol}). New results can be
expected from BOOMERanG\footnote{{\tt http://cmb.phys.cwru.edu/boomerang}},
MAXIPOL\footnote{{\tt http://groups.physics.umn.edu/cosmology/maxipol}} and
other experiments and later from
Planck\footnote{{\tt http://astro.estec.esa.nl/Planck}}.
For high frequency CMB measurements the most important foreground is
certainly the emission from Galactic Interstellar Dust
(ISD). Submillimetre and millimetre (hereafter submm) emission {\sl
intensity} of ISD can be inferred from IRAS and COBE--DIRBE
extrapolations ({\sl e.g.} \cite{sfd}) and has been measured on large
scales by COBE--FIRAS (\cite{firas1, boulanger, firas2}). On the other
hand, nothing is known on ISD {\sl polarization} in emission on scales
larger than 10~arcmin., {\sl i.e.} those precise scales which are the
most relevant for CMB studies. It is likely that ISD polarized
emission is the major foreground for high frequency CMB polarization
measurements. Ground--based observations of submm ISD polarization are
concentrated on high angular resolution (arcminute scale) of star
formation regions.  {\sl Indirect} evidence for large scale
polarization come from the polarization of starlight in absorption
(\cite{fosalba}). Goodman (1996) gives a review of the measurements
and ambiguities in the interpretation of the background starlight
polarization. In particular, the visible data are biased by low column
density lines of sight and do not fairly sample more heavily reddened
ones. {\sl Direct} submm measurements are therefore highly required
both for Galactic studies of the large scale coherence of the magnetic
field and in the field of CMB polarization, but are rather challenging
as they require sensitivities comparable to those of CMB studies.

Archeops\footnote{{\tt http://www.archeops.org}.} is an experiment
designed to obtain a large sky coverage in a single balloon
flight. First results on CMB anisotropies power spectrum are reported
in (\cite{benoit_cl, benoit_params}). Here, we present the first
results on ISD polarization measurements with Archeops. Its large sky
coverage strategy is optimized for finding fairly strongly polarized
sources without any bias on their location.

Section~\ref{se:desc} briefly describes the instrument and
Sect.~\ref{se:ground_calib} the ground based calibrations on polarized
channels. Section \ref{se:process} presents the specific processing
applied to the polarized data. More specifically
subsect.~\ref{se:inversion} presents the inversion method applied to
determine the Stokes parameters. Section~\ref{se:results} is dedicated
to the main results on local clouds and diffuse
regions. Section~\ref{se:syste} assesses the reliability of the
results and Sect.\ref{se:discussion} their physical interpretation.

\section{Description of the instrument \label{se:desc}}
\begin{figure}[!ht]
{\includegraphics[clip, angle=0, scale = 0.3]{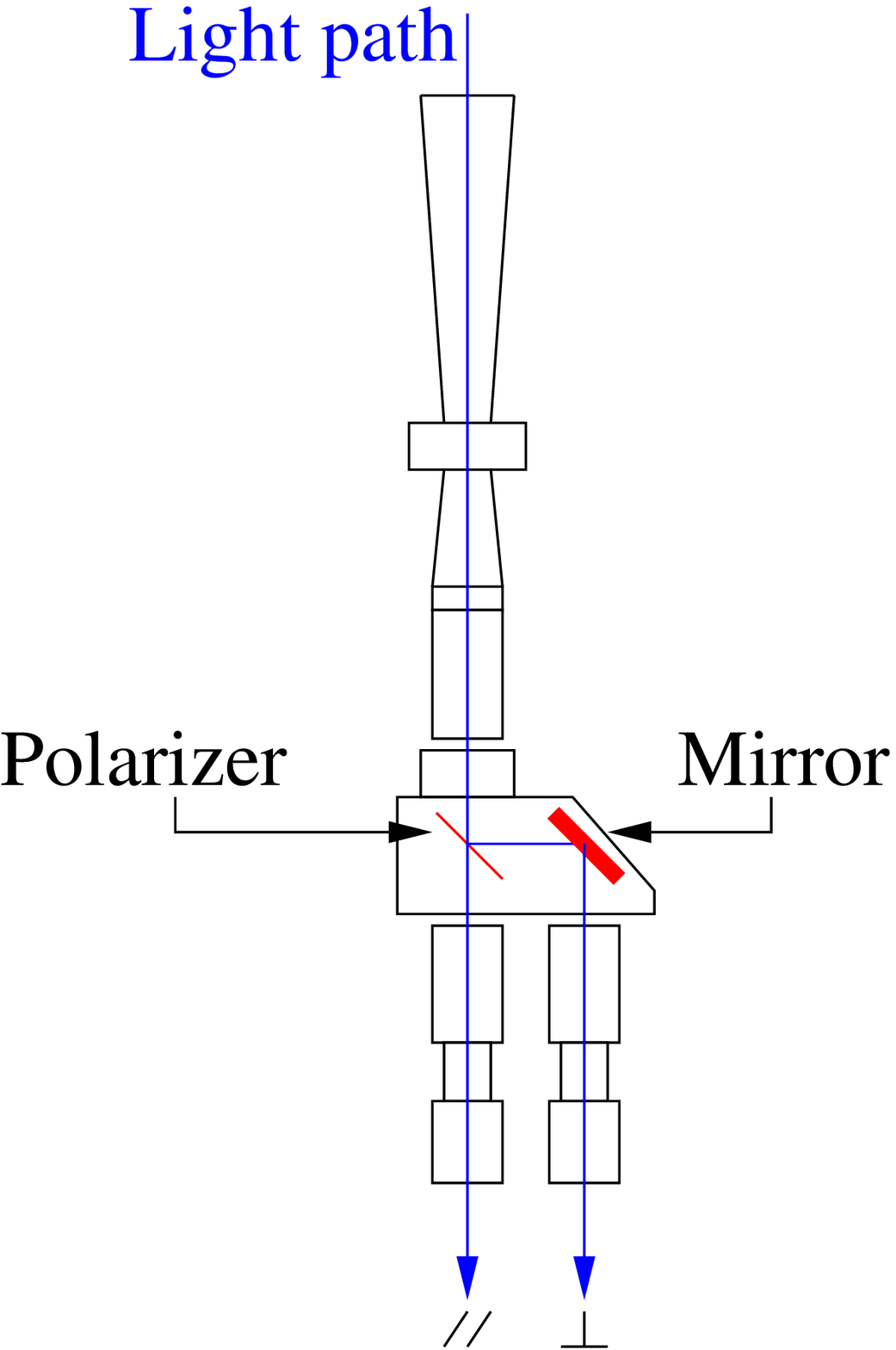}}
{\includegraphics[clip, angle=0, scale = 0.3]{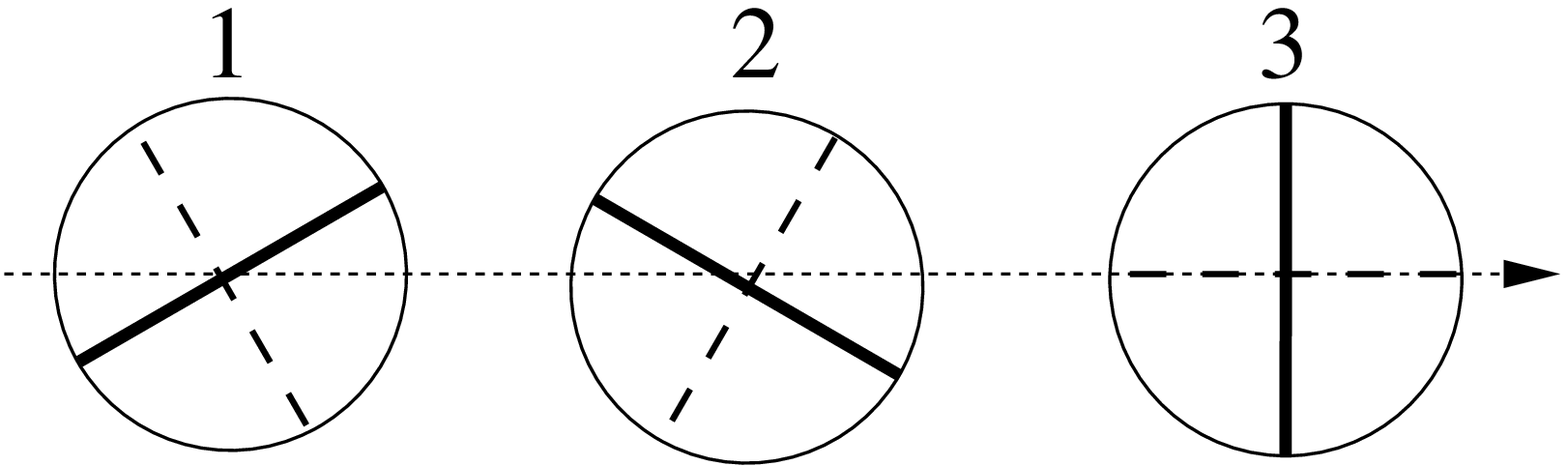}}
\caption{
a) Scheme of an OMT at 353~GHz used for Archeops. The back to back
horn is heat sunk to the 10~K stage. The box containing the polarizer
beam splitter and the mirror is on the 1.6~K stage. The bolometers on
the 100~mK stage and their associated horns are not shown. The light
enters from the top of the drawing into the back to back horn (one
horn per pair of bolometers). One polarization mode is transmitted
through the polarizer beam splitter to the first bolometer (A,
$\parallel$), the second mode is reflected to the second bolometer (B,
$\perp$). This system ensures that each bolometer of the same pair
sees the same point of the sky at the same time.  b) Orientation of
the polarization for each bolometer of the three OMTs in the focal
plane. The arrow represents the scan direction. The thick solid lines
refer to the $\parallel$ direction, the thick dashed lines refer to
the $\perp$ direction.}
\label{fig:omt}
\end{figure}

A detailed description of the instrument technical and inflight
performance is given in (\cite{benoit_app}, \cite{benoit_tech}); we
here provide only a summary description. Archeops is a balloon--borne
experiment with a 1.5~m off--axis Gregorian telescope described in
(\cite{Hanany:2002}). In particular, it satisfies the
Mizuguchi-Dragone condition (\cite{mizuguchi, dragone_82}) in which
there is negligible cross polarization at the center of the field of
view. The cryostat contains a bolometric array of 21 photometers
operating at frequency bands centered at 143~GHz (6~bolometers),
217~GHz (8), 353~GHz (6$~=$~3 polarized pairs) and 545~GHz (1). The
focal plane is maintained at a temperature of~$\sim 100$~mK using a
$^3$He-$^4$He dilution cryostat.  Observations are carried out by
rotating the payload at 2~rpm producing circular scans at a fixed
elevation of $\sim 41^\circ$. Pointing reconstruction is done \emph{a
posteriori} by using a dedicated optical stellar sensor made of a
40~cm optical telescope and 46 photodiodes. Observations of a single
night cover a large fraction of the sky as the circular scans drift
across the sky due to the rotation of the Earth. The experiment was
launched on February 7, 2002 by the CNES\footnote{Centre National
d'{\'E}tudes Spatiales, the French national space agency} from the
Swedish balloon base in Esrange, near Kiruna, Sweden, $68^\circ$N,
$20^\circ$E. It reached a float altitude of $\sim 34$~km and landed
21.5 hours later in Siberia near Noril'sk, where it was recovered by a
Franco-Russian team. The night--time scientific observations span
12~hours of integration from 15.0 UT to 3.0 UT the next day. The
polarized channels comprise of three quasi-optical modules, which are
equivalent to Ortho Mode Transducers (hereafter OMT, \cite{boifot},
\cite{chattopadhyay}). A pair of conjugated bolometers (see
Fig.~\ref{fig:omt}) is coupled to the sky through a single 10~K back
to back horn via the OMT. The OMT, which is attached to the 1.6~K
stage, is made from a single polarizing grid mounted at 45 degrees to
the horn axis to divide the incoming light into the two orthogonal
polarization modes. One is transmitted to the first bolometer (A), the
other is reflected towards the second one (B). At any time, the sum of
the two bolometer outputs measures the total intensity while the
difference measures the $Q$ Stokes parameters in the OMT eigen
basis\footnote{In this paper, the circular polarized mode $V$ is
assumed to be negligible and cannot be measured by our experimental
setup.}. The three OMT units are aligned along the scan direction and
have their (A)--polarization axis oriented at $60^\circ$ with respect
to each other in order to minimize errors in polarization
reconstruction (\cite{couchot}) (see~Fig.~\ref{fig:omt}).

\section{Ground--based calibration \label{se:ground_calib}}

Laboratory measurements were performed in order to calibrate the
transmission and orientations of the polarizers. The cross 
polarization of a single grid was measured to be less than 
1~\% and is neglected hereafter.

If $\tau$ is the intensity transmission rate and if $\parallel$
($\perp$) refers to the transmitting (extinguishing) direction of a
polarizer, then $K = (\tau_\parallel + \tau_\perp)/2$, $k =
(\tau_\parallel - \tau_\perp)/2$ and $q =
\sqrt{\tau_\parallel\tau_\perp}$ are the three parameters
characterizing a polarizer in Stokes formalism. For an ideal
polarizer, $K = 0.5$, $k = 0.5$, $q=0$. If ${\bf S} = (I, Q, U)$ is
defined with respect to the observation basis $(x, y)$ and describes
the polarization state of the radiation propagating along $-z$, and if
the $\parallel$ direction of the polarizer makes an angle $\alpha$
with $x$, then the transmitted Stokes vector is ${\bf S^\prime} =
{\cal M}{\bf S}$, with the Mueller matrix being

{\footnotesize
$$
{\cal M} = \left(\begin{array}{ccc}
K & k\cos2\alpha & k\sin2\alpha \\
k\cos2\alpha & K \cos^2 2\alpha + q\sin^2 2\alpha & (K-q)\cos2\alpha\sin2\alpha \\
k\sin2\alpha & (K-q)\cos2\alpha\sin2\alpha & K \sin^2 2\alpha + q\cos^2 2\alpha 
\end{array}\right).
$$
}

In case of an unpolarized incoming radiation $I_0$, a photometer
placed behind a polarizer receives $KI_0$. When it is placed behind
two polarizers that are oriented at angles $\alpha$ and $\varphi$, it receives
$(K_1K_2 + k_1k_2\cos2\alpha\cos2\varphi +
k_1k_2\sin2\alpha\sin2\varphi)I_0$.  In the case of the 353 GHz
bolometers, the OMT polarizer is fixed in the focal plane with an
angle $\varphi$. Rotating a calibration polarizer (hereafter CP) in
front of it and fitting the measured intensity as a function of $\alpha$
gives $k_1k_2\cos2\varphi$ and $k_1k_2\sin2\varphi$ from which
$\varphi$ can be deduced.

We place a box containing a calibration polarizer which rotates at
1.5~rpm above the entrance window of the cryostat. The is covered with
eccosorb to avoid parasitic reflections. It has two apertures. In one
aperture we place a 77~K thermal source made of polystyrene cup filled
with liquid nitrogen. The bottom of the cup is lined with
eccosorb. The other aperture faces the cryostat. The black body
emission is chopped at 13.4~Hz against the ambient temperature to
enable lock-in detection. This set up allows us to determine the
position of the grid polarizers in the OMT (see~Fig.~\ref{fig:omt}) to
within 3 degrees. This source of error contributes an uncertainty of
less than 5~\% in $Q$ and $U$.

During the ground based preparation of the flight, we placed a matrix
of $4\times 4$ grids of $50~\mu$m Cu/Be wires with a step of
$100~\mu$m in front of the $1\;\mathrm{m}^2$ 2~Hz modulated thermal
source (\cite{benoit_app}) placed on a hill at $\simeq 1$~km from the
telescope. This provided a linearly polarized blackbody source for an
additional pre-flight polarization calibration. We verified that the
orientation of the grid polarizers in the OMT agreed with the
laboratory measurements and found that the beam shape in the $Q$ and
$U$ states agreed with the $I$ beam shape within 20\%.

\section{Polarization data processing \label{se:process}}

\subsection{Standard processing \label{sse:standard_processing}}

The Stokes parameters reconstruction, as well as the
cross--calibration described above, only apply to clean data
associated with an accurate pointing. We here summarize the
preparation of the data described in more details in
(\cite{pipeline_paper}).

Pointing reconstruction is performed with about 200 detected stars
per revolution and provides an rms pointing accuracy better than 1.5
arcmin. The polarizers angles determined from ground calibrations
(see Sect.~\ref{se:ground_calib}) can then be computed on the sky.

The raw Time Ordered Information (TOI), sampled at $\simeq 153$~Hz,
are decompressed, then filtered to take into account the AC biasing
scheme coming from the readout electronics.  Cosmic rays, electronic
spikes, artifacts and noisy data are detected and flagged with an
automatic algorithm followed by visual inspection. Small areas around
strong point sources are flagged as well. The flagged data
representing less than 1.5\% of the data are replaced by a constrained
realisation of noise for subsequent detrending and high--pass
filtering.  The data are corrected with a bolometer model for drifts
of the instrument response due to changes in the background optical
loading and in the focal plane temperature.  Low frequency drifts due
to airmass and temperature fluctuations of the various stages of the
cryostat (0.1~K, 1.6~K, 10~K) are decorrelated using housekeeping data
(altitude, elevation, temperatures). A spin--synchronous
atmospheric signal remains and prevents us from using the Cosmological
Dipole for an accurate calibration. In--flight observations of Jupiter lead to the
determination of the bolometer time constant (used to deconvolve the
data stream) as well as the photometric pixel beams (with an error
less than 10~\%). The beams are identical within polarizer pairs and
moderately elliptical, with a minor and major axis FWHM of resp.~10.6
and 13.4~arcmin. We assume that in--flight $Q,\;U$ beams are identical
to the intensity beam.

\subsection{Filtering \label{filter}}

We briefly describe here the specific post--processing applied to the
353 GHz channels. This processing is not specific to polarization but
rather to Galactic studies for experiments that have 
a scan strategy like Archeops or Planck.

The major noise component that remains after the pipeline (as
described above) is some low frequency noise. A brute force low pass
filter applied on the timeline generates ringing on both sides of the
Galactic plane. The key issue is thus to remove the best low frequency
baseline without producing a significant ringing. To do this, we first
mask the Galaxy using a SFD template (\cite{sfd}) and use localized
slowly varying functions (\cite{benoit_tech}) to interpolate the
Galactic plane and obtain a first estimation of the baseline. This
estimation is used to perform a noise constrained realization of the
timeline on the masked area. Then, an optimized low frequency baseline is
calculated using wavelet shrinkage techniques (\cite{juan}) which allow to
remove high frequency noise. This baseline is subtracted from the
original timeline.

\subsection{Cross--calibration method \label{se:xcal_section}}

\begin{figure}[!ht]
\resizebox{\hsize}{!}
{\includegraphics[clip, angle=0]{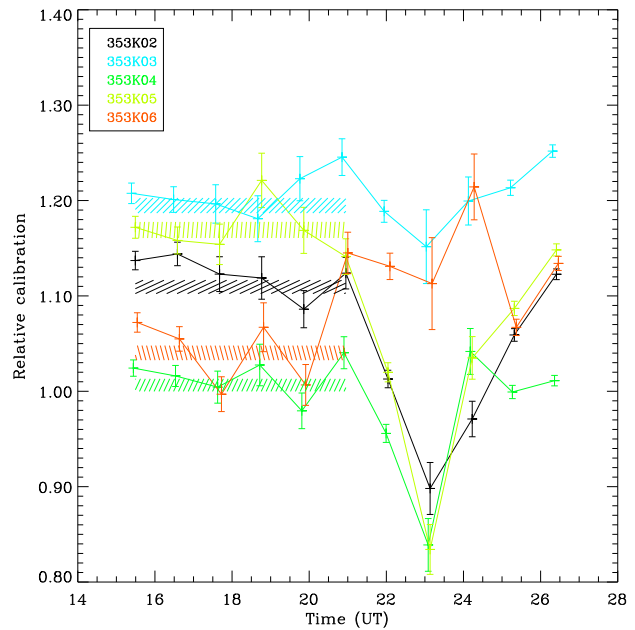}}
\caption{Variation of the cross--calibration coefficients $\alpha$ during the
flight relative to the reference baolometer 353K01. The coefficients
are mostly constant at the beginning of the flight (within error
bars), but become noisier after UT=21h30. The shaded areas show the
$\pm 1 \sigma$ values of the coefficients used in the present
analysis, and the time interval over which they are computed.}
\label{fig:evol_inter}
\end{figure}

Since polarization is obtained from differences of measurements from
detectors at various orientations, it is critical that they all be
accurately cross--calibrated. Any mismatch in this cross--calibration
automatically generates intensity leaks into the fainter polarization
mode. We found that the absolute calibrations obtained on Jupiter and
on the Galaxy (Sect.~\ref{calib}) are not precise enough for
polarization measurements: typically, in order to detect a 5\%
polarization on the Galaxy it is necessary to have a
cross--calibration accuracy of better than 2\%. To achieve higher
accuracy cross-calibration we have derived a method based on
inter--comparing the large signal coming from Galactic ``latitude
profiles'' from different bolometers. A latitude profile is a
tabulation of intensity as a function of latitude where the data at
all longitudes is averaged to produce a single intensity value in each
2 degrees latitude bin. Because latitude profiles average the
intensity from all longitudes they have a larger signal to noise ratio
compared to two dimensional maps of the galaxy. Also, it is plausible
to assume that for the latitude profiles a spatially uncorrelated
galactic polarization signal averages to zero. This last hypothesis
is very important and various tests to prove its validity are
discussed in Sect.~\ref{se:robustness_xcal}.

Let $s_1(b), \ldots, s_{n_{bol}}(b)$ be the $n_{bol}$ Galactic profiles ($b$ is
the latitude bin), measured by $n_{bol}$ bolometers. We make the
assumption that all these profiles are identical up to a calibration
factor $\alpha_j$ :

\begin{equation}
\label{eqn:model}
s_j(b) = \alpha_j\overline{s}(b)+n_j(b),
\end{equation}
with $\overline{s}(b)$ a reference profile and $n$ the noise. Making
the assumption of Gaussian white noise, we minimize the associated
$\chi^2$ with respect to the $\{\alpha_j\}_{j=1, n_{bol}}$ and
$\overline{s}$ simultaneously under the constraint that $\alpha_1 =
1$. The value of $\alpha_1$ is determined using the absolute
calibration. We have verified that the cross calibration does not
depend on which of the $\{\alpha_j\}$ is the constraining parameter.

Once calculated, these coefficients are used to compute $Q$ and $U$
maps (see Sect.~\ref{se:inversion}) and to check for the presence of a
polarized signal. If a residual polarized signal is detected in some
areas of the sky, we remove these areas before making the profiles,
and recompute the $\{\alpha_j\}_{j=1, n_{bol}}$. We have performed
simulations showing that with these two steps, the correct relative
calibration coefficients are recovered with a precision of better than
2\%. We also checked that the choice of the reference bolometer is
irrelevant.  To perform such simulations we used Galactic templates
provided by an extrapolation at 353~GHz of COBE and IRAS data
(\cite{sfd}, hereafter SFD) and included noise and the polarization
properties of the instrument. 

\begin{figure*}[!ht]
{\includegraphics[clip, angle=90, scale=0.5]{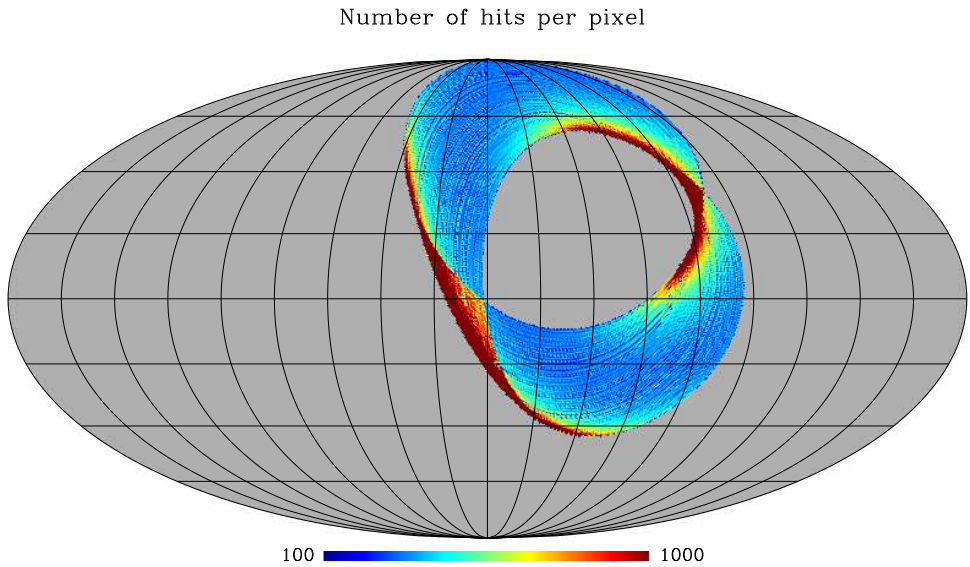}}
\caption{Total number of detector samples per pixels with the 6 bolometers at
353GHz. The map is centered on the Galactic anti--center and grid
coordinates are spaced by $20^\circ$. The Galactic plane is scanned
from about 85 to 120 and 180 to $200^\circ$. The covered fraction of
the sky is 17~\%.}
\label{fig:redundancy}
\end{figure*}

We calculate the cross-calibration factors using the entire data and
using only periods of 60 minutes and plot their evolution during the
flight in Fig.~\ref{fig:evol_inter}. The variance in the
cross-calibration increases starting at about 21hr UT. Around this
time the scans become more tangent to the Galactic plane. We attribute
the larger variance to the pattern of the scan and to noise induced by
the atmosphere. Simulations of the scans and a $1/f$ noise model
partially reproduce the larger variance; the simulations are limited
in their capability to simulate the actual noise arising from the
atmosphere. For the analysis in this paper we keep only data from
15h30 to 21h and Fig.~\ref{fig:evol_inter} shows the value of the
cross-calibration factors and their standard deviations that are used
for the analysis in this paper. The redundancy map corresponding to
this sky coverage is presented on Fig.~\ref{fig:redundancy}. The way
uncertainties in the cross calibration affect the degree of
polarization is discussed in Sect.~\ref{se:syste}.

\subsection{Absolute Calibration \label{calib}}

For absolute calibration we use the so-called FIRAS ``Dust spectrum
Maps'' data\footnote{{\tt http://space.gsfc.nasa.gov/astro/cobe/}},
which are spectral sky maps (from 100~$\mu$m to 4~mm for each $7^\circ$
pixel) from which the CMB, interplanetary dust and interstellar line
emission have been subtracted. Each FIRAS spectrum is fitted with a
modified blackbody emission law: $S =
\tau(\nu/\nu_0)^\beta P(\nu, T_{dust})$, where $\beta$ is an empirical
spectral index. This model is then convolved with the Archeops
bandpass filter. The Archeops data are smoothed to match FIRAS
beam. The calibration is then obtained from a correlation between
FIRAS and Archeops Galactic latitude profiles, and has an absolute
accuracy of about 6\%. This affects only the absolute values of
$I,\;Q,\;U$ and neither the degree of polarization nor its
orientation. A detailed description of the calibration is given in
(\cite{guilaine_paper,benoit_tech}). In order not to include
polarization effects in this calibration, we calibrated the total
intensity of each pair of cross--calibrated bolometers against
FIRAS. All numerical values throughout the paper are given in
$\mathrm{mK_{RJ}}$. A brightness of $1\;\mathrm{mK_{RJ}}$ is equivalent to
$4.36 \;\mathrm{MJy.sr}^{-1}$ using IRAS convention (constant $\nu
I_{\nu}$) for Archeops 353~GHz bandpass filter, and
15.4~$\mathrm{mK_{CMB}}$.

\subsection{Inversion method \label{se:inversion}}

For a given direction of observation ${\bf n}$, the associated usual
coordinate vectors $({\bf e_{\varphi}}, -{\bf e_{\theta}})$ tangential
to the sphere are chosen as the reference frame to express Stokes
parameters $(I, Q, U)$. The sign accounts of Galactic longitudes and
latitudes on the sphere that are clockwise oriented. Let ${\bf E}$ be
the incident electric field, $E$ its amplitude and $\psi$ its angle
with respect to ${\bf e_{\varphi}}$ in the tangential plane, defined
in the range $[0^{\circ}, 180^{\circ}]$, then

\begin{eqnarray}
Q & \equiv & |{\bf E} \cdot {\bf e_{\varphi}}|^2 - |{\bf E} \cdot {\bf e_{\theta}}|^2 \\
U & \equiv & |{\bf E} \cdot {\bf e_{\varphi}^{45}}|^2 - |{\bf E} \cdot {\bf e_{\theta}^{45}}|^2,
\end{eqnarray}
where the superscript 45 means that the original coordinate vectors have been
rotated clockwise by 45 degrees. 

In this subsection, the polarizers
are assumed to be calibrated against an unpolarized source
(cf. subsec.~\ref{se:xcal_section}, ~\ref{calib}). The {\sl
calibrated} polarimeter at an angle $\alpha$ with respect to ${\bf
e_{\varphi}}$ measures

\begin{eqnarray}
m(\alpha) & = & c\;E^2\cos^2(\alpha - \psi) + n \nonumber \\
          & = & c\;(I + Q\cos2\alpha + U\sin2\alpha) + n,
\label{m_i}
\end{eqnarray}
where the noise $n$ depends on time, $\alpha$ depends on the bolometer
and on the pixel, and $c$ is the bolometer calibration coefficient. In
order to make a $n_{pix}$ map, all samples must be taken into account
to include noise correlations (in time and from pixel to pixel) and
equation (\ref{m_i}) is generalized to :

\begin{equation}
{\bf M} = \mathcal{A}{\bf S} + {\bf N},
\label{stokes_system}
\end{equation}
where {\bf M} is the time ordered vector of the $n_t \times n_{bol}$
measures, {\bf S} the $(3~n_{pix})$--vector Stokes map of the sky,
$\mathcal{A}$ the pointing matrix encoding the pointing
information and polarizer angles and {\bf N} the $n_t \times n_{bol}$
noise vector. The $\chi^2$ is given by

\begin{equation}
\chi^2 = (\bf{M} - \mathcal{A}{\bf S})^T \mathcal{N}^{-1}({\bf M} - \mathcal{A}{\bf S})
\label{chi_2}
\end{equation}
and is minimized by the solution

\begin{equation}
{\bf S} = (\mathcal{A}^T\mathcal{N}^{-1}\mathcal{A})^{-1}
\mathcal{A}^T\mathcal{N}^{-1}{\bf M}.
\label{inv_sys}
\end{equation}
The covariance matrix is then 

\begin{equation}
{\bf \Sigma} = (\mathcal{A}^T\mathcal{N}^{-1}\mathcal{A})^{-1}.
\label{cov_mat}
\end{equation}
Solving the linear system (\ref{inv_sys}) is one of the recurrent
problems in CMB studies since the matrices and vectors are usually
large. In our case, however, the size of the polarized regions
correspond to temporal frequencies where the noise is essentially
white (in--scan induced noise), and the level of striping in $Q$ and
$U$ (cross--scan induced noise) is negligible and therefore we can use
the following simplification. When the noise is not correlated from
one measurement to another, $\mathcal{N}$ is diagonal and the
inversion of large matrices can be avoided. We therefore consider each
pixel individually, compute the (3,3)-matrix
$\mathcal{A}^T\mathcal{N}^{-1}\mathcal{A}$ and the (3)-vector
$\mathcal{A}^T\mathcal{N}^{-1}{\bf M}$. The system of equations thus
involves small mathematical objects and the inversion time is small.

\section{Results \label{se:results}}

\begin{figure*}[!ht]
{\includegraphics[clip, angle=90, scale=0.5]{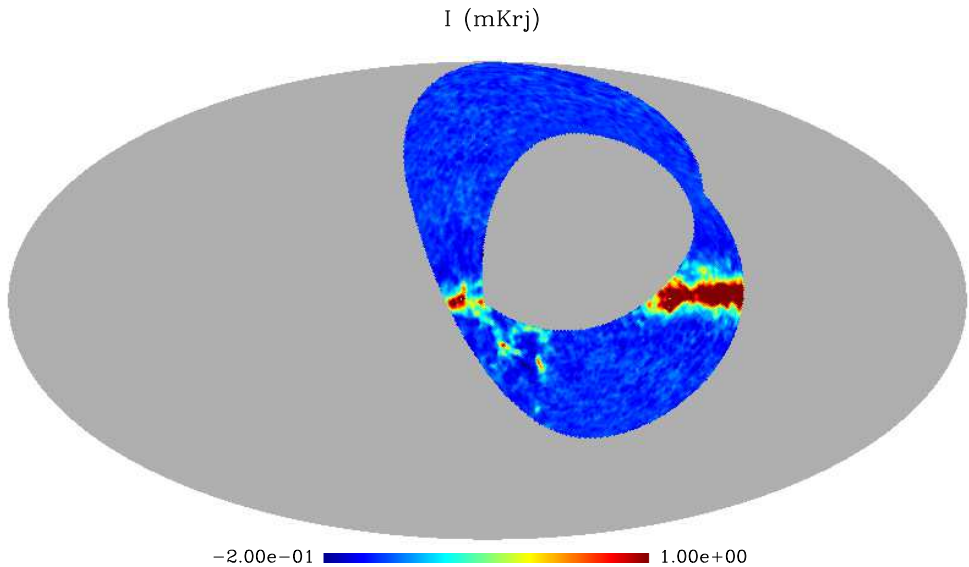}}
\caption{
Archeops $I$ map at 353 GHz in $\mathrm{mK}_{RJ}$ smoothed with a 1 degree
Gaussian beam.}
\label{fig:i_map}
\end{figure*}

\begin{figure*}[!ht]
{\includegraphics[clip, angle=90, scale=0.5]{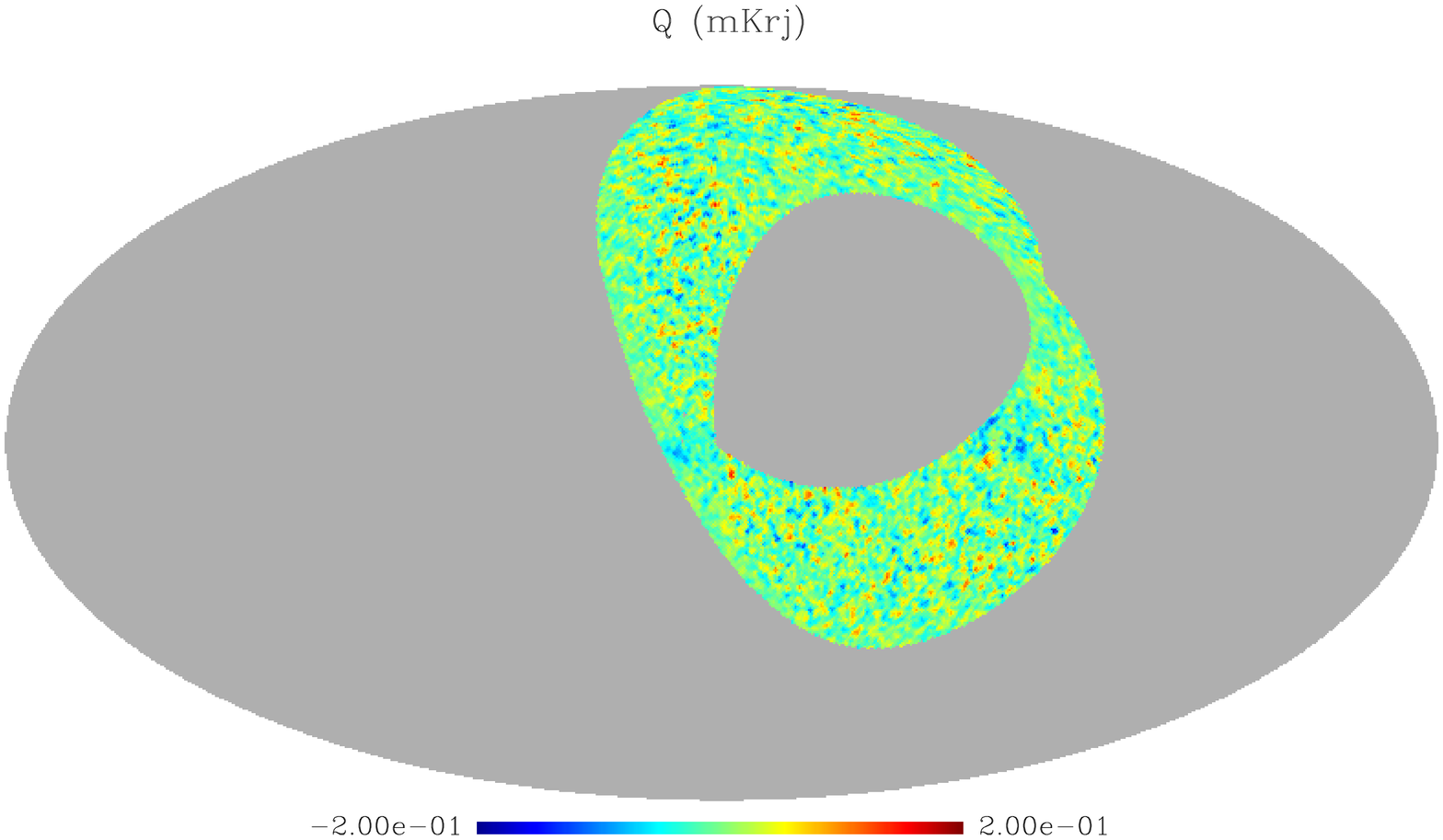}}
\caption{
Archeops $Q$ map at 353 GHz in $\mathrm{mK}_{RJ}$ smoothed with a 1 degree
Gaussian beam.}
\label{fig:q_map}
\end{figure*}

\begin{figure*}[!ht]
{\includegraphics[clip, angle=90, scale=0.5]{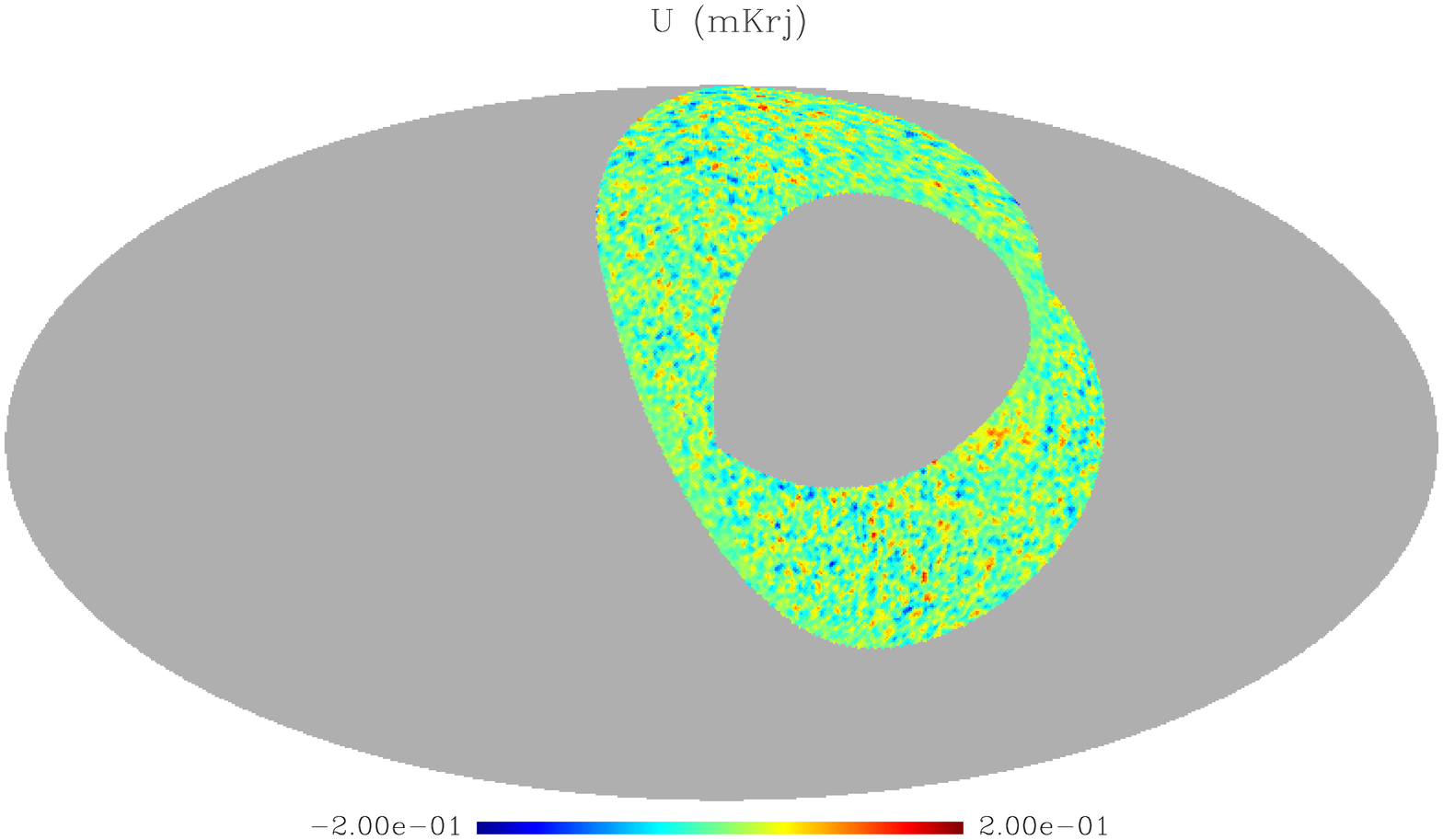}}
\caption{
Archeops $U$ map at 353 GHz in $\mathrm{mK}_{RJ}$ smoothed with a 1 degree
Gaussian beam.}
\label{fig:u_map}
\end{figure*}

Once the data are cleaned (Sect.~\ref{sse:standard_processing}), they
are filtered (Sect.~\ref{filter}), calibrated (Sect.~\ref{calib}) and
combined according to Eq.~(\ref{stokes_system}) and inverted with
Eq.~(\ref{inv_sys}) to produce maps of $I$, $Q$, and $U$. We choose a
pixel size of 27.5~arcmin, corresponding to HEALpix (\cite{healpix})
resolution parameter $n_{\mathrm{side}} = 128$. Pixels that have less
than 100 detector samples, which correspond to 0.11~sec mission
integration time and a $1\,\sigma$ $I$ noise level of
$143~\mu\mathrm{K_{RJ}}$ are blanked.  For display purposes the maps
are smoothed with a 1~deg beam, and these maps are shown in
Figs.~\ref{fig:i_map}~\ref{fig:q_map}~\ref{fig:u_map}).  The noise
estimate for $I, Q, U$ is obtained through Eq.~\ref{cov_mat}.

\begin{figure*}[!ht]
{\includegraphics[clip, angle=90, scale=0.5]{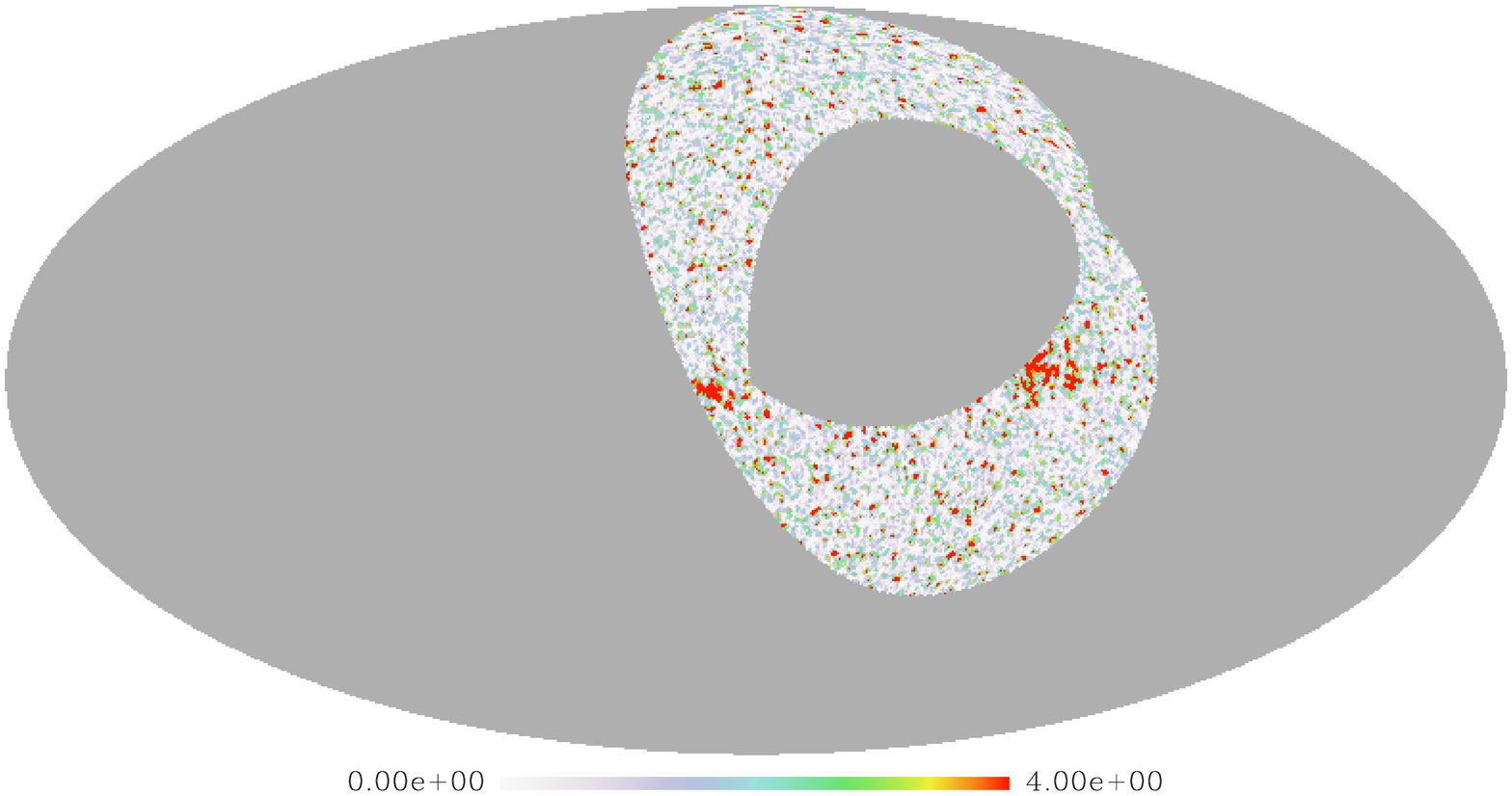}}
\caption{ Map of the normalized squared polarized intensity $(Q^2 +
U^2)/(\sigma_Q^2 + \sigma_U^2)$.
Twice this quantity is statistically distributed like a $\chi^2$ with
2 degrees of freedom. The 68, 95.4, 99.7\% CL of the mapped quantity
correspond to 1.1, 3.1, 5.8 respectively.}
\label{fig:map_ipol}
\end{figure*}

The dispersion of $Q$ and $U$ at high Galactic latitudes is found to
be $\simeq 1.1$ times larger than their noise estimates from the
inversion method. This is due to the fact that the noise is
not perfectly white. In the analysis that follows we use the measured dispersion
as a measure of the noise and not the lower noise estimated from 
the inversion method. The
instantaneous mission $I$ sensitivity is found to be about
48~$\mu\mathrm{K_{RJ}.sec}^{1/2}$. On average, the 1~$\sigma$ noise
per pixel of 27~arcmin ($n_{\mathrm{side}} = 128$) is found to be
82~$\mu\mathrm{K_{RJ}}$ in intensity and 105~$\mu\mathrm{K_{RJ}}$ in
$Q$ and $U$. A statistically significant polarization signal is
detected in various locations on the galactic plane.
Fig.~\ref{fig:map_ipol} shows a map of the ``normalized squared
polarized intensity'' defined as the squared polarized intensity $(Q^2
+ U^2)$ normalized to its variance $(\sigma_Q^2 + \sigma_U^2)$. Twice
this quantity behaves has a $\chi^{2}$ probability distribution
function with 2 degrees of freedom.
A statistically significant signal appears in regions where
the normalized squared polarized intensity significantly exceeds a
value of unity and several such regions are detected along the
galactic plane. We now discuss the results on isolated regions and
diffuse medium.

\subsection{Galactic dense clouds \label{ss:dense_clouds}}

Here we focus on connected regions in which the normalized polarized
intensity exceeds the $2\sigma$ level on Fig.~\ref{fig:map_ipol}. The
Stokes parameters for these clouds are determined by averaging the
pixel values of the unsmoothed maps with weights that are inversely
proportional to the variance in the
pixels. Table~\ref{tab:results_clouds} gives the values of $I$, $Q$,
$U$, $p$ and $\theta$ for these regions. Because the Taurus cloud
region has been well studied by various instruments we give its values
separately to facilitate an easy comparison. Since $p$ is not a
Gaussian variable, we estimate and correct for the bias on its
determination using Monte Carlo simulations. Error bars on $p$ and
$\theta$ at 68~\% CL are also determined using simulations. The
simulations include the cross-calibration errors discussed in
Section~\ref{se:xcal_section}.

\renewcommand{\arraystretch}{1.6}
\begin{table*}[!h]
\begin{center}
\begin{tabular}{c|rrrr}
\hline
Cloud index & $I\;(\mathrm{mK_{RJ}})$ (stat) (syst)& $Q\;(\mathrm{mK_{RJ}})$ (stat) (syst)& $U\;(\mathrm{mK_{RJ}})$ (stat) (syst)\\
\hline
\hline
0 & $  1.018\pm  0.015 \pm  0.085 $&$ -0.057\pm  0.021 \pm  0.001 $&$  0.111\pm  0.017 \pm  0.008 $ \\
\hline
1 & $  0.562\pm  0.014 \pm  0.090 $&$ -0.115\pm  0.019 \pm  0.001 $&$  0.052\pm  0.016 \pm  0.003 $ \\
\hline
2 & $  1.498\pm  0.011 \pm  0.105 $&$ -0.004\pm  0.014 \pm  0.007 $&$  0.113\pm  0.013 \pm  0.014 $ \\
\hline
3 & $  0.419\pm  0.021 \pm  0.111 $&$  0.100\pm  0.028 \pm  0.013 $&$ -0.018\pm  0.027 \pm  0.009 $ \\
\hline
4 & $  0.994\pm  0.023 \pm  0.075 $&$ -0.125\pm  0.029 \pm  0.015 $&$  0.015\pm  0.029 \pm  0.010 $ \\
\hline
5 & $  0.820\pm  0.011 \pm  0.113 $&$ -0.135\pm  0.014 \pm  0.010 $&$  0.005\pm  0.015 \pm  0.005 $ \\
\hline
6 & $  0.698\pm  0.004 \pm  0.055 $&$ -0.059\pm  0.005 \pm  0.013 $&$  0.011\pm  0.006 \pm  0.004 $ \\
\hline
\hline
7 & $  0.409\pm  0.010 \pm  0.039 $&$ -0.023\pm  0.014 \pm  0.002 $&$ -0.010\pm  0.013 \pm  0.005 $ \\
\hline
8 & $  0.271\pm  0.006 \pm  0.065 $&$ -0.001\pm  0.009 \pm  0.002 $&$ -0.021\pm  0.008 \pm  0.006 $ \\
\hline
9 & $  0.473\pm  0.009 \pm  0.080 $&$  0.001\pm  0.013 \pm  0.003 $&$  0.001\pm  0.011 \pm  0.008 $ \\
\hline
\end{tabular}
\begin{tabular}{c|rrr|rr}
\hline
Cloud index & $l$ & $b$ & Size ($\mathrm{deg}^2$) & $p\;(\%)$ (stat) (syst)& $\theta\;(^\circ)$ (stat) (syst)\\
\hline
\hline
0 & 103.0 &   1.8 &   5.9 & $   12.1^{+   1.8}_{-   1.8} \pm    1.8 $&$   59\pm    4.7 \pm    1.0 $ \\
\hline
1 & 105.8 &   0.6 &   7.3 & $   22.2^{+   3.4}_{-   3.3} \pm    4.0 $&$   78\pm    3.7 \pm    0.8 $ \\
\hline
2 & 109.7 &   2.1 &   8.8 & $    7.5^{+   0.9}_{-   0.9} \pm    1.5 $&$   46\pm    3.6 \pm    2.0 $ \\
\hline
3 & 113.2 &  -2.7 &   2.9 & $   23.3^{+   6.5}_{-   6.7} \pm    9.7 $&$  175\pm    7.5 \pm    3.2 $ \\
\hline
4 & 113.6 &  -1.2 &   2.3 & $   12.3^{+   2.8}_{-   2.9} \pm    2.6 $&$   87\pm    6.7 \pm    2.8 $ \\
\hline
5 & 115.0 &   2.4 &   5.9 & $   16.3^{+   1.7}_{-   1.7} \pm    3.5 $&$   89\pm    3.2 \pm    1.1 $ \\
\hline
6 & 193.0 &   0.0 &  21.6 & $    8.5^{+   0.7}_{-   0.7} \pm    2.6 $&$   85\pm    2.8 \pm    3.1 $ \\
\hline
7 & 159.3 & -20.1 &  18.5 & $    5.3^{+   3.1}_{-   3.1} \pm    1.5 $&$  101\pm   13.6 \pm    6.3 $ \\
\hline
8 & 165.6 &  -9.0 &  50.1 & $    7.2^{+   2.8}_{-   2.8} \pm    4.1 $&$  133\pm   11.5 \pm    3.8 $ \\
\hline
9 & 174.4 & -13.6 &  21.0 & $   < 3.4 $&$   23\pm   16.0 \pm  104.7 $ \\
\hline
\hline
\end{tabular}
\caption{
\emph{Top:} Stokes parameters of significantly polarized Galactic
clouds (above double line) and of Taurus complex (below double line).
\emph{Bottom: } 
Coordinates of the maximum of intensity, measured area, degree and
orientation of polarization of these clouds. Systematic error bars are
derived from the dispersion of the results with different filtering
parameters on the timelines (see Sect.~\ref{se:filtering}). Last line
correspond to a 95~\%~CL upper limit. Angles are counted clockwise, 0
being parallel to the Galactic plane. A brightness of
$1\;\mathrm{mK_{RJ}}$ is equivalent to $4.36
\;\mathrm{MJy.sr}^{-1}$ using IRAS convention (constant $\nu
I_{\nu}$) for Archeops 353~GHz bandpass filter. Absolute
calibration error of 6~\% is not included.}
\label{tab:results_clouds}
\end{center}
\end{table*}

\subsection{Diffuse Galactic regions}

\begin{figure*}[!ht]
\begin{center}
{\includegraphics[clip, angle=0, scale=0.4]{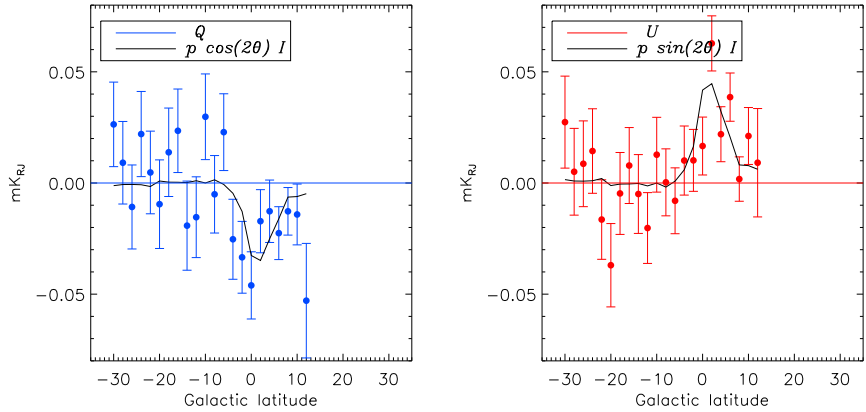}}
\caption{ Scaling of $I,\;Q,\;U$ Galactic latitude profiles
{\sl e.~g.}~ for a Galactic longitude range $[105, 110]^\circ$. The profiles
enable to constrain diffuse polarization.}
\label{fig:diffuse_profile}
\end{center}
\end{figure*}

\begin{figure*}[!ht]
\resizebox{\hsize}{!}
{\includegraphics[clip, angle=0, scale=0.3]{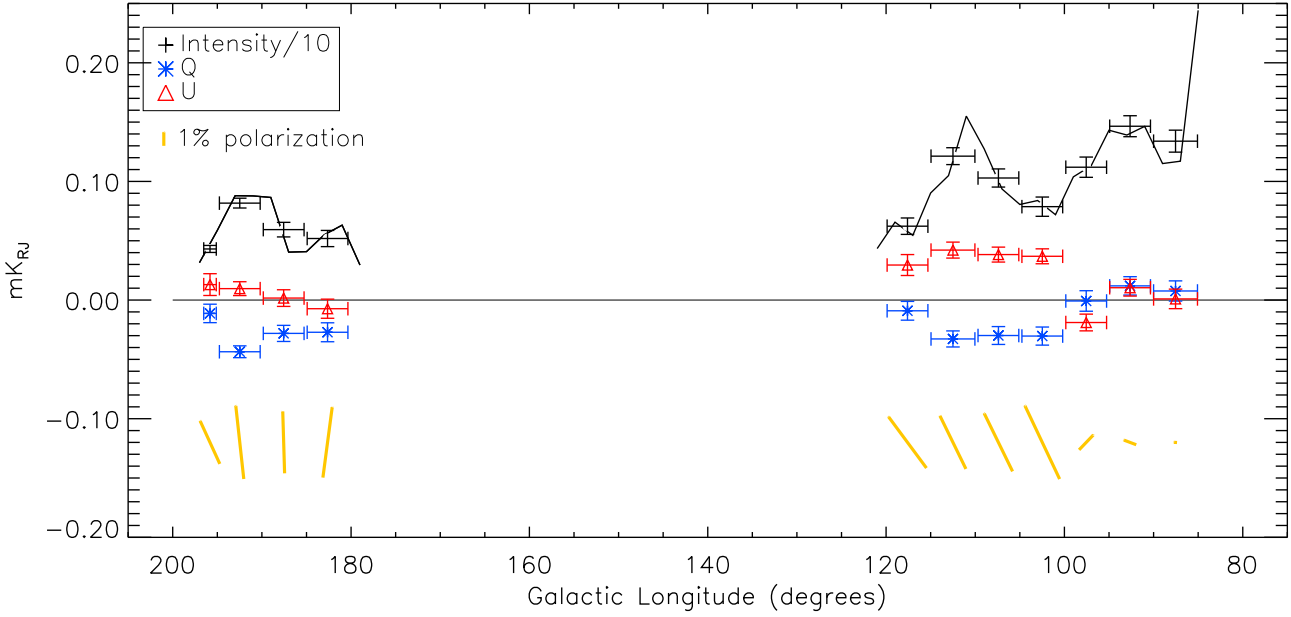}}
\caption{ Summary figure of diffuse Galactic polarization. The
intensity (divided by 10) is represented in black and is taken to be
the average value for $-2 \leq b \leq 2$ in each longitude band. The
thin solid line is the same value in $2^\circ$ wide bands. The
direction of polarization for every bin is represented below in green,
and the length of the dash is proportionnal to the degree of
polarization $p$ in \%. The horizontal error bars represent the width
of the longitude bins, which is $5^\circ$ except for the edge bins.
Values are summarized in Tab.~\ref{tab:diffuse}.}
\label{fig:plot_synthese}
\end{figure*}

In this section we determine whether there is a coherent level of
polarization on large regions without defining any cloud
boundaries. For that purpose and to enhance the signal to noise ratio,
we divide the Galaxy into 5~deg wide bands along Galactic
longitude. For each band we construct three latitude 
profiles consisting of the values of $I$, $Q$ and $U$ as a function 
of latitude; we use binning of 2 degrees in latitude. These three profiles are
then used to find a unique polarization vector $(p, \theta)$
characterizing the region corresponding to the profile. We avoid the
bias in the determination of the polarization vector using
simulations. An example of a profile is shown in
Fig.~\ref{fig:diffuse_profile}, and the results are summarized in
Table~\ref{tab:diffuse} and in Fig.~\ref{fig:plot_synthese}. Coherent
polarization levels of few percents are significantly detected up to
5~\% at the 3 to 4~$\sigma$ level for several longitude bands, some of
which include the clouds already discussed in the previous
section. Even after masking these clouds, a significant coherent
polarization remains in the same longitude bands.

\renewcommand{\arraystretch}{1.6}
\begin{table*}[!h]
\begin{center}
\begin{tabular}{c|rrrr}
\hline
Gal. Long. Range $(^\circ)$ & $I\;(\mathrm{mK_{RJ}})$ (stat) (syst)& $Q\;(\mathrm{mK_{RJ}})$ (stat) (syst)& $U\;(\mathrm{mK_{RJ}})$ (stat) (syst)\\
\hline
$     85 \;\;     90 $&$   1.34 \pm  0.093 \pm  0.040 $&$  0.008 \pm  0.009 \pm  0.000 $&$  0.001 \pm  0.008 \pm  0.004 $\\
$     90 \;\;     95 $&$   1.47 \pm  0.088 \pm  0.053 $&$  0.012 \pm  0.008 \pm  0.002 $&$  0.010 \pm  0.007 \pm  0.002 $\\
$     95 \;\;    100 $&$   1.12 \pm  0.085 \pm  0.052 $&$ -0.001 \pm  0.008 \pm  0.000 $&$ -0.019 \pm  0.007 \pm  0.003 $\\
$    100 \;\;    105 $&$   0.79 \pm  0.081 \pm  0.062 $&$ -0.030 \pm  0.008 \pm  0.001 $&$  0.037 \pm  0.006 \pm  0.005 $\\
$    105 \;\;    110 $&$   1.03 \pm  0.076 \pm  0.062 $&$ -0.030 \pm  0.007 \pm  0.003 $&$  0.038 \pm  0.006 \pm  0.001 $\\
$    110 \;\;    115 $&$   1.21 \pm  0.071 \pm  0.031 $&$ -0.033 \pm  0.007 \pm  0.000 $&$  0.042 \pm  0.007 \pm  0.002 $\\
$    115 \;\;    120 $&$   0.62 \pm  0.069 \pm  0.029 $&$ -0.009 \pm  0.008 \pm  0.009 $&$  0.030 \pm  0.009 \pm  0.006 $\\
$    180 \;\;    185 $&$   0.52 \pm  0.068 \pm  0.048 $&$ -0.027 \pm  0.008 \pm  0.011 $&$ -0.007 \pm  0.008 \pm  0.000 $\\
$    185 \;\;    190 $&$   0.59 \pm  0.061 \pm  0.055 $&$ -0.028 \pm  0.006 \pm  0.008 $&$  0.002 \pm  0.007 \pm  0.004 $\\
$    190 \;\;    195 $&$   0.82 \pm  0.041 \pm  0.055 $&$ -0.044 \pm  0.005 \pm  0.011 $&$  0.010 \pm  0.006 \pm  0.001 $\\
$    195 \;\;    197 $&$   0.43 \pm  0.029 \pm  0.042 $&$ -0.011 \pm  0.008 \pm  0.005 $&$  0.013 \pm  0.009 \pm  0.003 $\\
\hline
\end{tabular}

\begin{tabular}{r|rr}
\hline
Gal. Long. Range $(^\circ)$ & $p\;(\%)$ (stat) (syst)& $\theta\;(^\circ)$ (stat) (syst)\\
\hline
$     85 \;\;     90 $&$    0.2^{+   0.6}_{-   0.5} \pm    0.1 $&$    4 \pm 34 \pm 15 $ \\
$     90 \;\;     95 $&$    0.9^{+   0.5}_{-   0.5} \pm    0.3 $&$   20 \pm 16 \pm  6 $ \\
$     95 \;\;    100 $&$    1.6^{+   0.7}_{-   0.6} \pm    0.4 $&$  134 \pm 15 \pm  0 $ \\
$    100 \;\;    105 $&$    6.0^{+   0.9}_{-   0.8} \pm    1.0 $&$   65 \pm  4 \pm  2 $ \\
$    105 \;\;    110 $&$    4.7^{+   0.6}_{-   0.7} \pm    0.6 $&$   64 \pm  4 \pm  1 $ \\
$    110 \;\;    115 $&$    4.3^{+   0.5}_{-   0.6} \pm    0.2 $&$   64 \pm  4 \pm  0 $ \\
$    115 \;\;    120 $&$    4.8^{+   1.3}_{-   1.5} \pm    1.6 $&$   54 \pm  8 \pm  9 $ \\
$    180 \;\;    185 $&$    5.2^{+   1.6}_{-   1.4} \pm    2.6 $&$   98 \pm  9 \pm  3 $ \\
$    185 \;\;    190 $&$    4.6^{+   1.0}_{-   1.0} \pm    1.9 $&$   88 \pm  8 \pm  4 $ \\
$    190 \;\;    195 $&$    5.4^{+   0.6}_{-   0.6} \pm    1.7 $&$   84 \pm  4 \pm  1 $ \\
$    195 \;\;    197 $&$    3.5^{+   1.9}_{-   2.1} \pm    1.6 $&$   65 \pm 17 \pm  8 $ \\
\hline
\end{tabular}
\caption{
\emph{Top: }Stokes parameters of $5^\circ$ Galactic longitude 
bands. The intensity is the average over the longitude band and $b$
taken in the range $[-2^\circ, 2^\circ]$.  $Q$ and $U$ are scaled from
that intensity using the latitude profile fit.
\emph{Bottom: } Degree and orientation of polarization for these bands. 
Angles are counted clockwise, 0 being parallel to the Galactic plane.
Due to incomplete sky coverage some longitude bands are not quoted.
Systematic error bars are derived from the dispersion of the results
with different filtering parameters on the timelines
(Sect.~\ref{se:filtering}).}
\label{tab:diffuse}
\end{center}
\end{table*}

\section{Systematics and cross--checks \label{se:syste}}

It is the first time that a significant detection of polarization
from the diffuse submillimetre Galactic dust emission is
reported. Before giving interpretation, we discuss the levels of
possible systematics that can alter the results.

\subsection{Cross checking different methods of polarization
determination}

To test our $Q$ and $U$ results, we have employed two additional
techniques to derive their values. Instead of finding a combined
solution for $I$, $Q$, and $U$ using equation (\ref{inv_sys}) we
determined $(Q, U)$ from differences of cross--calibrated pair of
bolometers by differencing the time-ordered data. Following equation
(\ref{m_i}) one can write

\begin{eqnarray}
\Delta m(\alpha) & \equiv & m(\alpha)/c - m(\alpha + \pi/2)/c^\prime \nonumber \\
                 & = & 2\;(Q\cos2\alpha + U\sin2\alpha) + n^\prime. \label{d_eq}
\end{eqnarray}

where $c$ and $c^\prime$ are the calibration constants of the
bolometers (see Eq.~\ref{m_i}). In this method the information on the
total intensity $I$ is lost.  However, the noise power spectrum of the
difference $\Delta m$ is much flatter at low frequencies than for
individual bolometers because the differencing scheme removes common
spurious unpolarized signals such as the atmosphere or common gain
drifts.

We apply the map making algorithm outlined in
eqs.~(\ref{stokes_system}, \ref{inv_sys}) to the difference $\Delta m$
and re--derive $Q$ and $U$ values. The results are consistent within
one $\sigma$ with the results reported in Tables
\ref{tab:results_clouds} and \ref{tab:diffuse} for all clouds and galactic profiles.

\begin{figure}[!ht]
{\includegraphics[clip, scale=0.35, angle=0]{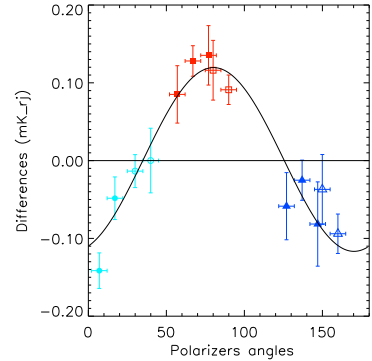}}
\caption{
Fit of difference--timelines (filtered differences of bolometer
outputs of the same pair) as a function of their associated polarizer
angle for the Gemini cloud (cloud index 6 in Table 1). This cloud is
observed at two different time intervals and provides useful
consistency check, independent of the inversion method. Values are
averaged in $10^\circ$ angle bins. Different symbols and colors are
for the different pairs Red squares, light blue circles and blue
triangles are for the first, second and third pair resp. Empty symbols
refer to the first time interval, filled ones refer to the second
one. The global fit has a $\chi^2/ndf$ of 0.93 and gives: $Q = -0.056
\pm 0.006~\mathrm{mK_{RJ}}$, $U = 0.020 \pm 0.006~\mathrm{mK_{RJ}}$
and $\theta = 80.2 \pm 3.4$~deg, in good agreement with values deduced
from the maps and mentioned in Tab.~\ref{tab:results_clouds}.}
\label{fig:phase_diagram}
\end{figure}

In the second technique we simply bin the difference of the TOD of a
given cloud in $10^{\circ}$ bins of the polarizer angle $\alpha$. The
binned signal is fit with a function of the form of Eq.~(\ref{d_eq})
to obtain $Q$ and $U$ values.  This method does not depend on the map
making algorithm which was used both as the main analysis technique
and for the alternate method described earlier in this Section. The
results of the binning for the Gemini cloud (centered on $(l,b) =
(194.5, -0.9)$) are shown in Fig.~\ref{fig:phase_diagram}, but have
been carried out for all the clouds for which
Table~\ref{tab:results_clouds} gives results.  For the Gemini cloud
the fit has a $\chi^2/ndf$ of 0.93 and gives $Q = -0.056 \pm
0.006~\mathrm{mK_{RJ}}$ , $U = 0.020 \pm 0.006~\mathrm{mK_{RJ}}$ and
$\theta = 80.2 \pm 3.4$~deg. The results show that Eq.~(\ref{d_eq}) is
a good fit to the binned data and that the $Q$ and $U$ values are in
good agreement with polarization values deduced from the other two
techniques described earlier. Figure~\ref{fig:phase_diagram} also
shows consistency between the two sets of three photometers (each with
two orthogonal bolometers) and between the data taken at two different
times during the flight.

\subsection{Consistency between bolometers}

\begin{figure}[!ht]
{\includegraphics[clip, scale=0.25]{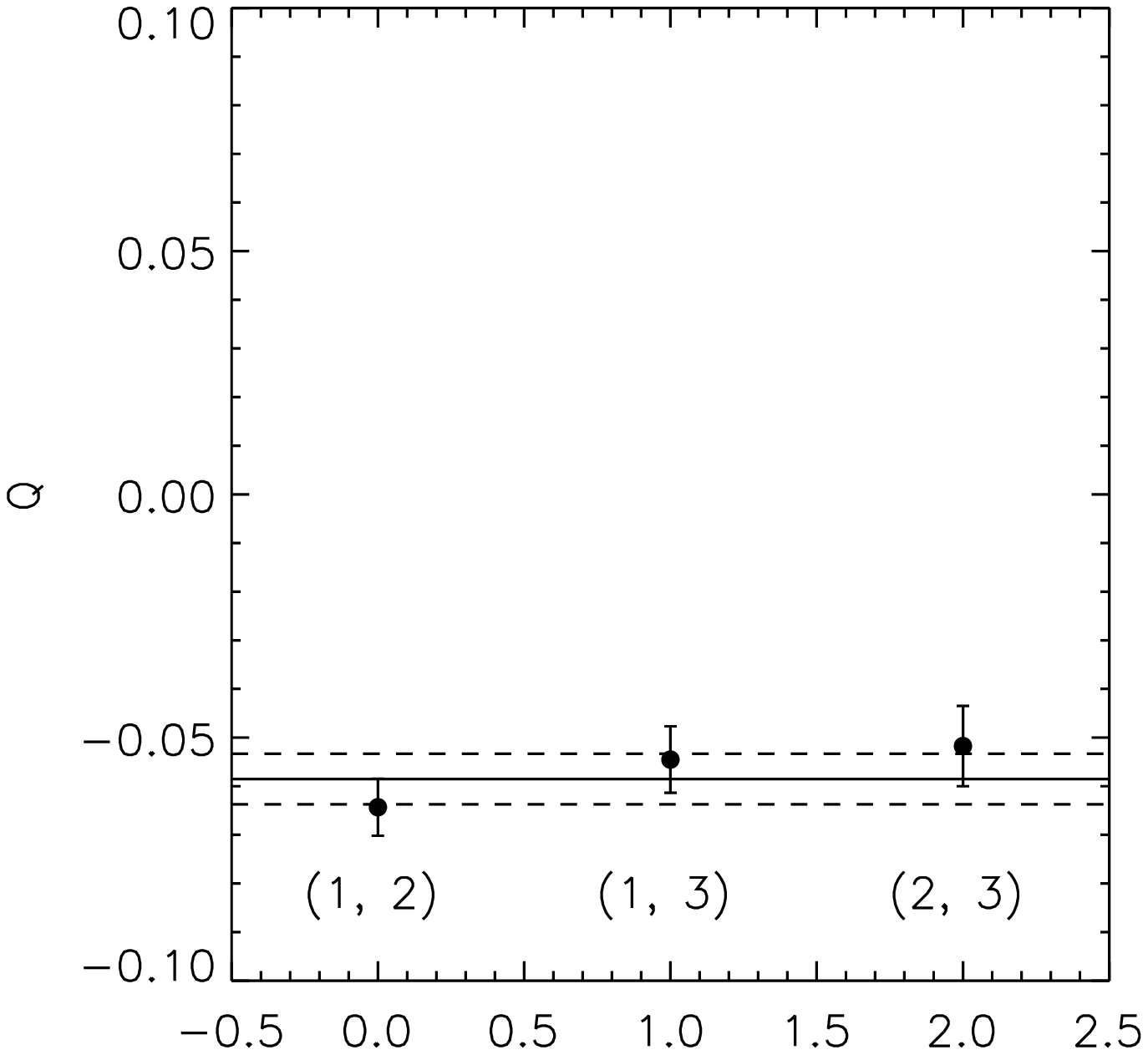}}
{\includegraphics[clip, scale=0.25]{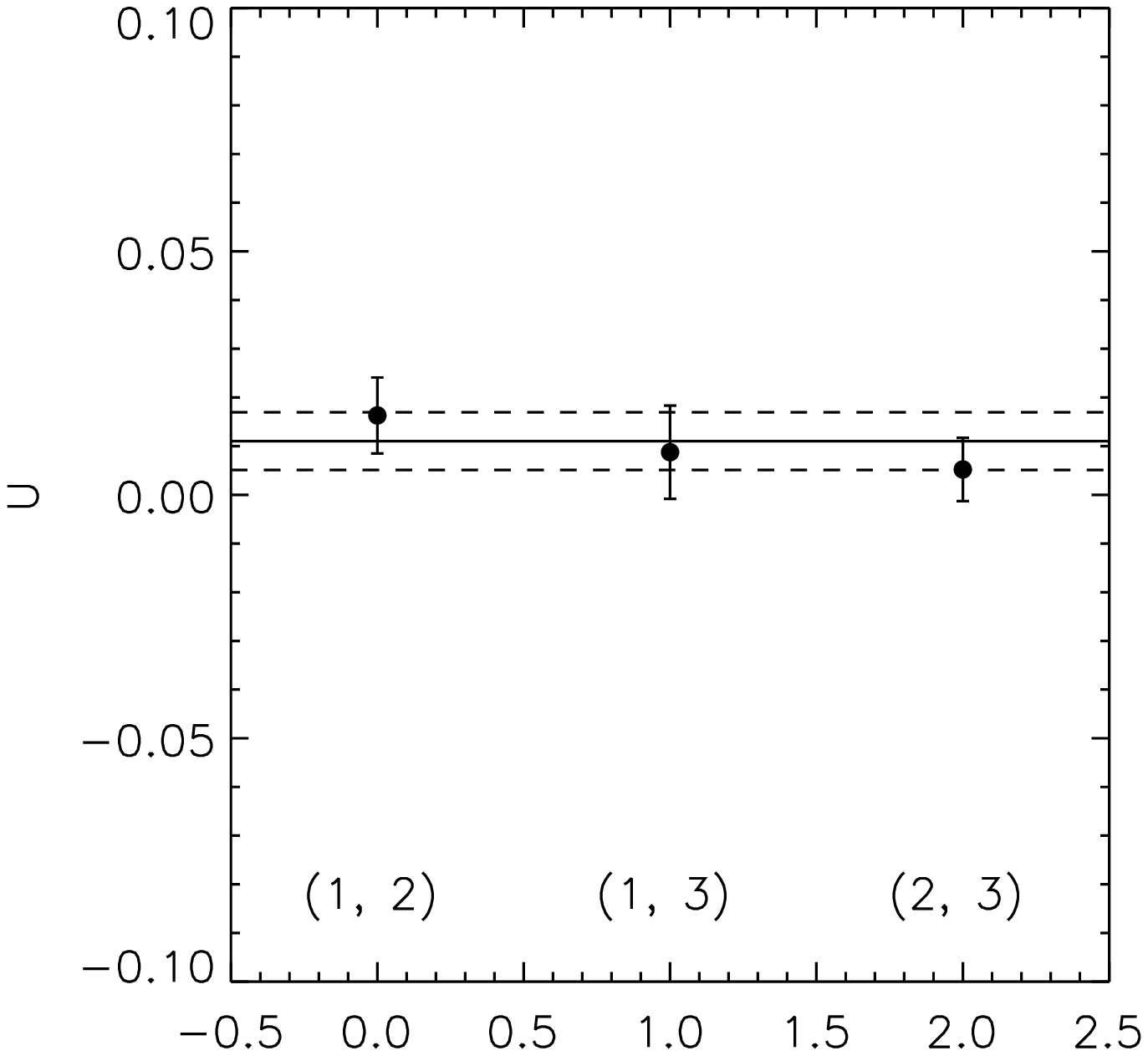}}
\caption{ $Q$ (left), $U$ (right) determined on the Gemini cloud 
($l \simeq 193^\circ,~b \simeq -1.2^\circ$) with sets of two pairs of 353~GHz bolometers
out of the three available. Each point is labeled with
the pairs used to compute it. The solid line is the value determined
on the map using the three pairs, the dashed lines are the 1 $\sigma$
errors on this value.}
\label{fig:two_pairs_consistency}
\end{figure}

In order to check the consistency between the three pairs of
bolometers, we computed $Q$ and $U$ with various combinations of only
two pairs out of the three
available. Figure~\ref{fig:two_pairs_consistency} shows that all the
results are in good agreement and are consistent with the values
derived using all three pairs. Moreover, the photometric accuracy of
the $I$ map can be checked against SFD template at 353~GHz. We find a
consistency within 10~\%.

\subsection{Uncertainties on cross--calibration \label{se:robustness_xcal}}

The validity of the cross-calibration procedure depends critically
on the assumption that the regions over which signals from different
detectors are compared are on average not polarized. This assumption
can fail in various ways and here we test for these failures.

If one region which is highly polarized biases the cross-calibration
coefficients, then the polarization derived in other locations in the
galaxy should be correlated with intensity. Visual inspection of the
maps does not reveal an obvious correlation. For example, the right
part of the intensity map (near Cygnus), which is the brightest
(Fig.~\ref{fig:i_map}), has no polarization counterpart. 

A large scale coherent polarization can induce a systematic error in
the cross-calibration if the polarizers crossed the Galaxy at
constant angles. Because of our scan pattern the polarizers rotate as
they cross the galaxy. To understand the effect quantitatively we
carried out simulations where the entire galaxy (\cite{sfd}) was
polarized at the 5\% level with constant orientation. The
reconstructed cross--calibration coefficients, derived using the
assumption that the galaxy does {\it not} have a large scale coherent
polarization, were biased at the 2-3\% level and modified the
polarization at the level $\Delta p \sim 1~\%$.

The galaxy is most probably not polarized at a constant level and
orientation. We derive the cross-calibration errors by simulating
several test maps of the Galaxy using SFD templates with a 5\%
polarization with random orientations in various regions. We perform a
first iteration of the cross calibration assuming that these simulated
galaxy maps are not polarized, and we then reconstruct the
polarization map using the derived cross-calibration coefficients. We
perform a second cross-calibration after masking the regions that were
found to be polarized after the first iteration with a significance
level of larger than 2 $\sigma$. The $1 \sigma$ uncertainty in the
extraction of the cross calibration coefficients after this second
iteration was 2\%. The $1 \sigma$ uncertainty in the determination of
the Stokes parameters in the regions that had 5\% polarization was
less than 1\% in $I$, and about 3--5~\% in $Q$ and $U$, depending on
the intensity of the chosen polarized region.

\begin{figure*}[!ht]
{\includegraphics[clip, scale=0.3]{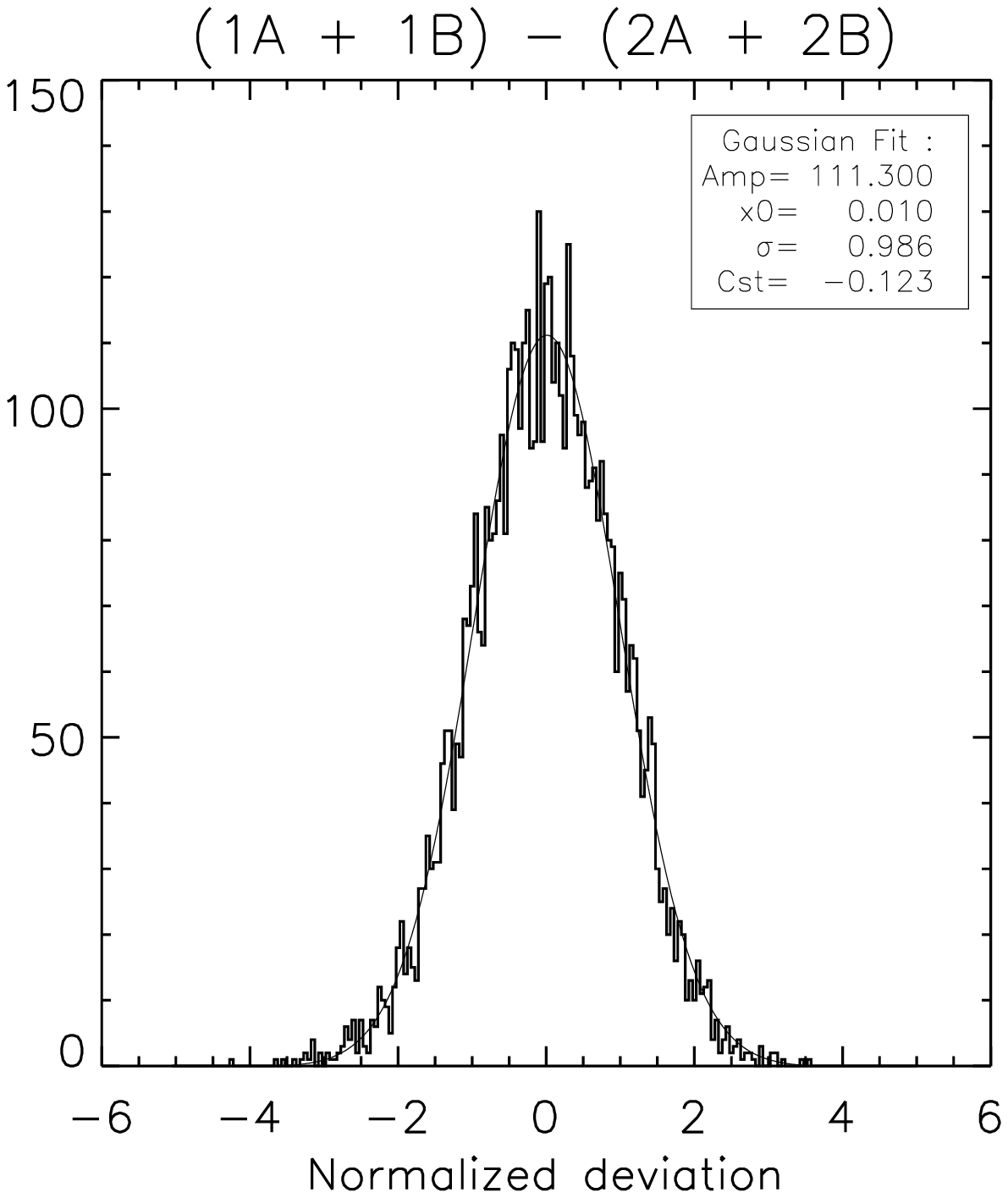}}
{\includegraphics[clip, scale=0.3]{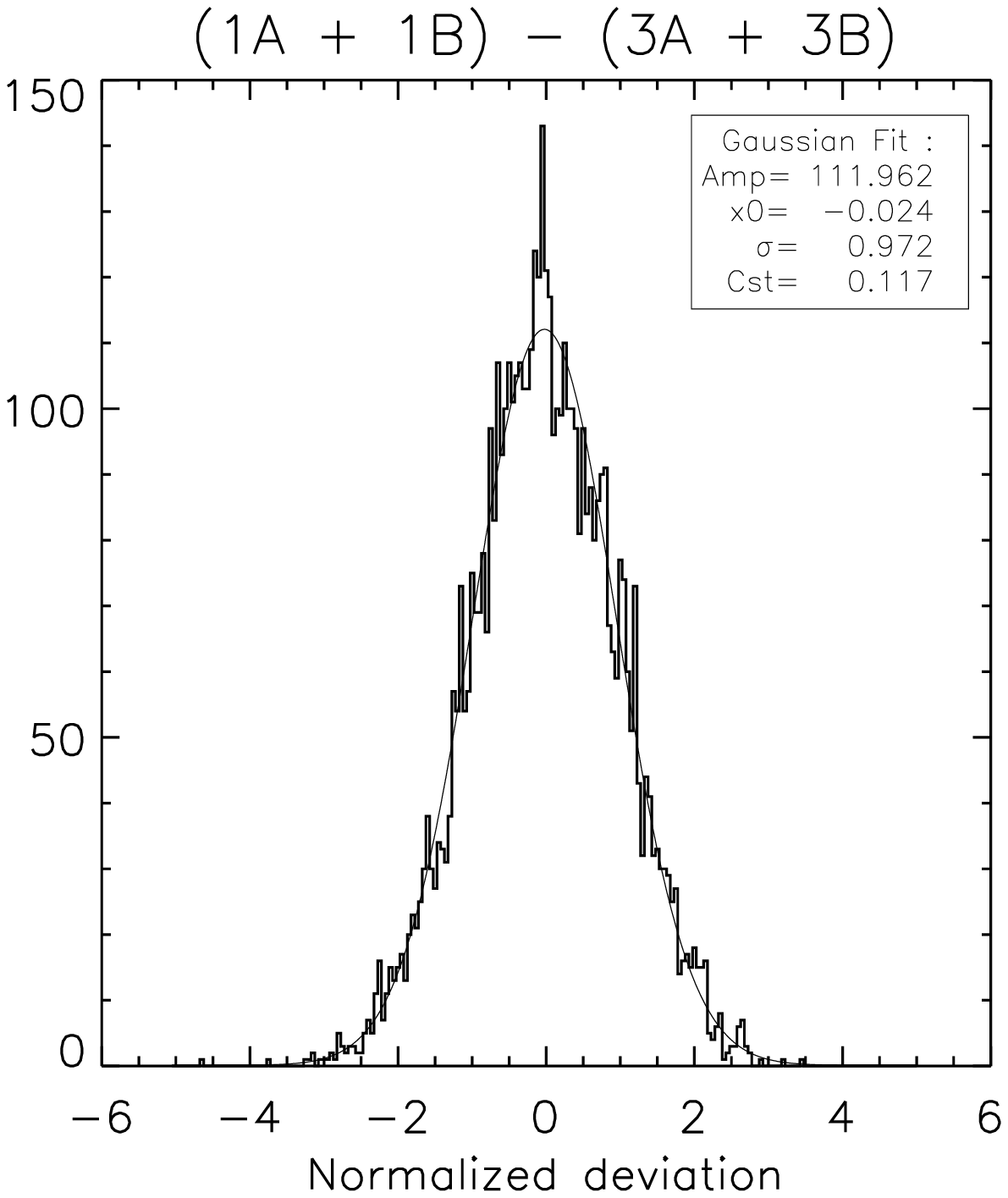}}
{\includegraphics[clip, scale=0.3]{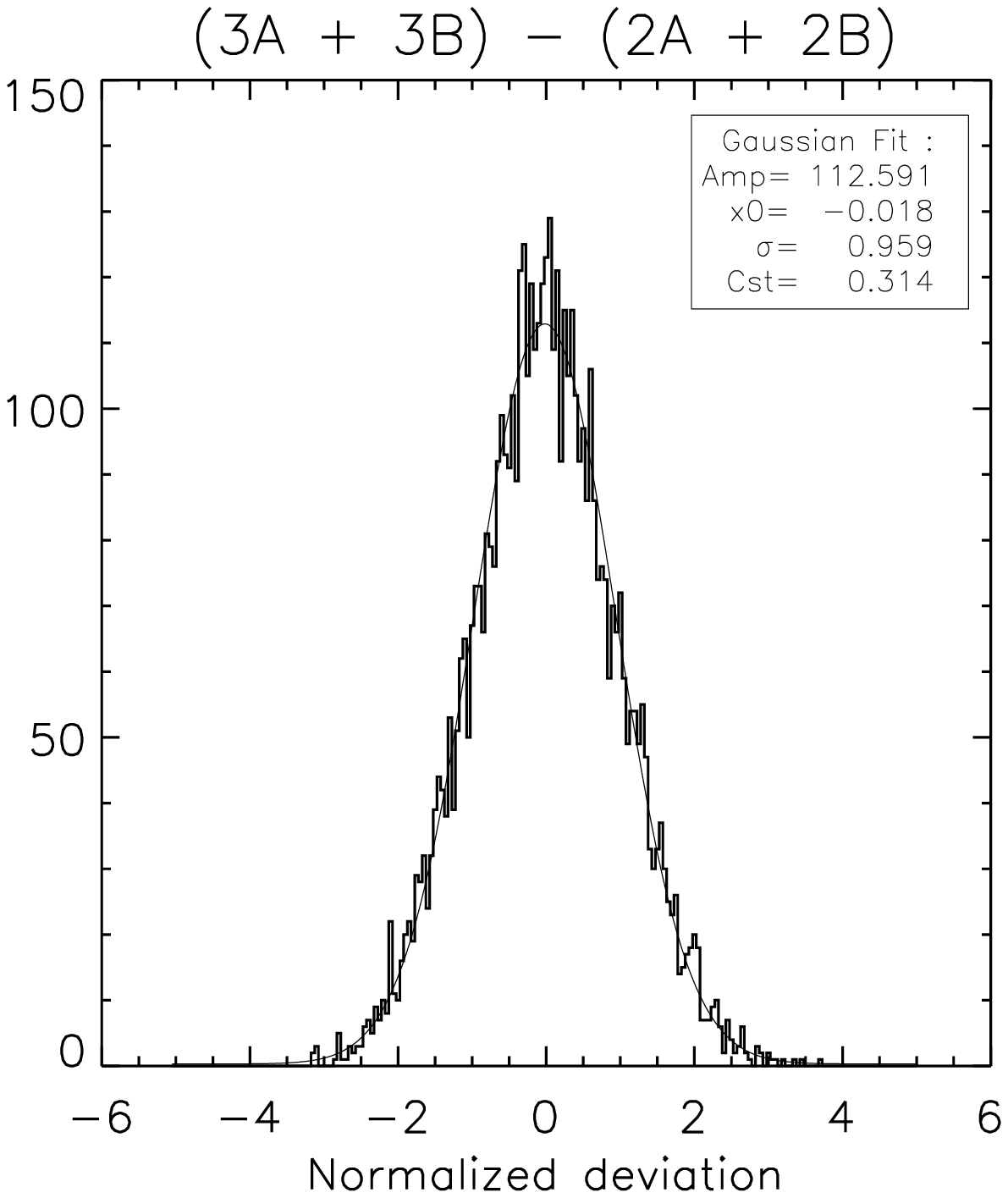}}
\caption{Histograms of the noise normalized difference between the 
intensity from two pairs of bolometers on the Galaxy (\mbox{$b \in [-10,
10]$}). These histograms are compatible with a normal distribution,
which shows a good relative calibration reconstruction.}
\label{fig:histo_horns_gal}
\end{figure*}

We can also check the cross--calibration process on the three intensity
maps that can be deduced from the three pairs of bolometers. The
histograms shown on Fig.~\ref{fig:histo_horns_gal} give the intensity
difference between two pairs normalized by the expected noise. The
histograms are Gaussian with a standard deviation of 1 and without any
outlyers.

In order to test the reconstruction of the noise characteristics, we
also perform a null test with the data themselves by randomizing the
angle of each polarizer pair at each sample before applying the
inversion (\ref{inv_sys}).  As maps are computed with rather large
pixels ($n_{side} = 128$), this effectively cancels out any diffuse
sky polarized signal as checked on simulations while conserving the
noise properties. If one considers the values of $Q$ and $U$ derived
from these ``randomized'' maps for the ten clouds of
Table~\ref{tab:results_clouds}, one can form the $\chi^2/ndf$ of the
hypothesis that they all be zero. We find a compatibility at the 90\%
CL.

\subsection{Filtering effects \label{se:filtering}}

Using the above mentioned simulations (\cite{sfd}), we observe that
time domain filtering removes the large scale diffuse emission (which broadly
has a cosecant law behaviour). It represents usually 10--20~\%,
sometimes 30~\% of the total intensity along the line of sight. By
changing the filtering parameters on the real data timelines (e.g. the
mask, the frequency cut), we observe a similar effect, principally on
$I$ and much less on $Q$ and $U$.
We derive the systematic error bars (Tabs.~\ref{tab:results_clouds}
and \ref{tab:diffuse}) from the dispersion of the results obtained
with these different filterings. The polarized emission characterized
by $Q$ and $U$ and its orientation are therefore more accurately
determined than the degree of polarization.

\subsection{Beam effects and time constants}

Beams are found to be nearly identical between the two bolometers of a
same pair. Nevertheless, slight beam mismatch convolved with the
Galactic gradient could generate a spurious polarization signal. This
effect has been estimated to be of at most 10~$\mu K$, below our
statistical uncertainties. This is also true for uncertainties on
time constants which are less than 2~millisec ({\sl i.e.} 
6~arcmin). The cross--calibration found using Jupiter  
agrees with the cross-calibration using the Galactic profiles 
to better than $1.5\;\sigma$ uncertainty for each bolometer.

\section{Interpretation of the results \label{se:discussion}}

The emission of two cloud complexes appear to be strongly polarized at
353~GHz. One large complex is in Cassiopeia (clouds 1--5 in
Tab.~\ref{tab:results_clouds}) with an area of
$33~\mathrm{deg}^2$. This area includes the supernova remnant CasA,
although the center is detected in the processing as a point source
and is not projected. The other complex coincides with the southern
part of Gem~OB1 (cloud 6 in
Tab.~\ref{tab:results_clouds}). Interestingly, the observed part of
the Cygnus complex is not fount to be significantly polarized.

The orientation of the polarization that we find using the galactic
profiles is found to be coherent on large scales and is also
consistent with that found in clouds. Overall, the orientation is
nearly orthogonal to the Galactic plane. It was long noted from
optical polarization studies that neighbouring stars had similar
polarization directions, with a degree of polarization $p_V$ more or
less proportional to reddening giving an empirical relation : $p_V =
0.03A_V$ (\cite{whittet, goodman}). The basic explanation is that a
large scale Galactic magnetic field induces alignment of elongated ISD
grains. Starlight polarization measures the projection of the
direction of polarization on the plane of the sky. However, optical
polarization measurements sample only rather low reddening lines of
sight and near infrared polarimetric studies yield ambiguous results
concerning denser clouds (\cite{whittet,goodman}). On the other hand,
submm polarization is free from opacity effects and samples all the
ISD material along the line of sight. The 353~GHz band is nearly on
the Rayleigh--Jeans side of dust thermal emission, so grains of
various temperatures should not have very different contributions in
different radiation fields. If the grains that produce visible
extinction are responsible for submm emission with the same
efficiency, then an average polarization of at least 3~\%
(\cite{stein}) is expected at 353~GHz. As shown in
Tab.~\ref{tab:diffuse}, we find a level slightly above this figure and
much higher in some clouds therefore indicating a very efficient grain
alignment mechanism (\cite{hilde_95}). Moreover, starlight extinction
polarization measurements are predicted to be orthogonal to the
polarized thermal emission. Catalogs of starlight polarization have
been gathered (\cite{fosalba} and references therein) and show a
global orientation parallel to the Galactic plane in this longitude
range, compatible within 20 to $30^\circ$ with the orientation of
diffuse medium emission as shown in Tab.~\ref{tab:diffuse} and
Fig.~\ref{fig:plot_synthese}. If the magnetic field follows the spiral
arms, one can also expect, as we measure only its projection onto the
plane of the sky, that some longitudes should have a reduced apparent
polarization (see Fig.~5 in \cite{fosalba}). The very low polarization
found on Cygnus is in qualitative agreement with this prediction as
the spiral arm lies along the line of sight in this longitude range.

A coherence of the orientation of polarization between the diffuse
medium and denser clouds is generally observed, except for the
cloud~G113.2-2.7. It seems that the global magnetic field that
pervades the Galactic plane also goes deeply into some denser clouds
and is not tangled by turbulence effects. However, the degree of
polarization may vary by as much as a factor two inside the same cloud
complex. This probably comes from the local variability of the
magnetic field.

The present observations are complementary to the far infrared and
millimetre polarimetry as reviewed by (\cite{hildebrand_96}) because
here we probe much more diffuse lines of sight.

Although the instrument sensitivity does not allow to measure directly
high Galactic latitude dust polarization, we can extrapolate our
results to these regions, assuming that the coherence of the magnetic
field and the properties of the ISD are similar to the ones in the
Galactic plane. It can then be anticipated that dust polarized
emission will be the major foreground to CMB polarization studies at
the level of 10~\% of its intensity, as anticipated by
(\cite{prunet}). The integration along the line of sight of various
orientations tends to decrease the overall effect of polarization in
the Galactic plane, whereas at high latitude, this depolarization
effect should be smaller.

\section{Conclusions}

Archeops provides the first large coverage maps of Galactic submm
emission with 13~arcmin resolution and polarimetric capabilities at
353~GHz. We find that the diffuse emission of the Galactic plane in
the observed longitude range is polarized at the 4-5~\% level except
in the vicinity of the Cygnus region. Its orientation is mostly
perpendicular to the Galactic plane and orthogonal, as expected, to
the orientation of starlight polarized extinction. Several clouds of
few square degrees appear to be polarized at more than 10~\%. This
suggests a powerful grain alignment mechanism throughout the
interstellar medium. Our findings are compatible with models where a
strong coherent magnetic field coplanar to the Galactic plane follows
the spiral arms, as observed in external Galaxies.

\begin{acknowledgements}
  The authors would like to thank the following institutions for funding
  and balloon launching capabilities: CNES (French space agency), PNC
  (French Cosmology Program), ASI (Italian Space Agency), PPARC, NASA,
  the University of Minnesota, the American Astronomical Society and a
  CMBNet Research Fellowship from the European Commission.  Healpix
  package was used throughout the data analysis~(\cite{healpix}).
\end{acknowledgements}


\end{document}